\documentclass[12pt]{article}
\usepackage{epsfig, epsf, graphicx}
\usepackage{pstricks, pst-node, psfrag}
\usepackage{amssymb,amsmath,bm}
\usepackage{verbatim,enumerate}
\usepackage{diagbox}
\usepackage{rotating, lscape}
\usepackage{setspace}
\usepackage{paralist}
\usepackage{apacite}
\usepackage{natbib}
\usepackage{color, colortbl}
\usepackage{tabu}
\usepackage{xcolor}
\usepackage{contour}
\usepackage{paralist}

\usepackage{algorithm}
\usepackage{algorithmic}

\usepackage{enumitem}

\usepackage{appendix}

\usepackage[labelformat=simple]{subcaption}

\DeclareCaptionLabelFormat{myformat}{\tablename\ #2}
\makeatletter
\renewcommand\p@subtable{} 

\usepackage[framemethod=default]{mdframed}
\usepackage{showexpl}

\mdfdefinestyle{exampledefault}{%
linecolor=black,linewidth=1.5pt,
middlelinewidth=3pt,rightline=true,innerleftmargin=10,innerrightmargin=10}
\usepackage{tikz}
\usepackage{multirow}
\usepackage{bbm}
\usetikzlibrary{arrows,calc,tikzmark,shapes,fit,positioning}
\tikzset{box/.style={draw, rectangle, thick, text centered, minimum height=3em}}
  \tikzset{line/.style={draw, thick, -latex'}}
\usepackage{hyperref}
\usepackage{booktabs}
\usepackage{threeparttable}
\usepackage{float}
\usepackage{multicol,listings,siunitx}
\usepackage{caption}
\def\boxit#1{\vbox{\hrule\hbox{\vrule\kern6pt
          \vbox{\kern6pt#1\kern6pt}\kern6pt\vrule}\hrule}}

\def\bse{\begin{eqnarray*}}
\def\ese{\end{eqnarray*}}
\def\be{\begin{eqnarray}}
\def\ee{\end{eqnarray}}
\def\bq{\begin{equation}}
\def\eq{\end{equation}}
\def\bse{\begin{eqnarray*}}
\def\ese{\end{eqnarray*}}

\definecolor{seagreen}{rgb}{0.18,0.55,0.34}
\definecolor{lawngreen}{rgb}{0.49,0.99,0}
\definecolor{lightsalmon}{rgb}{1,0.63,0.48}
\definecolor{lightyellow}{rgb}{0.99,0.906,0.429}

\DeclareFontFamily{OT1}{pzc}{}
\DeclareFontShape{OT1}{pzc}{m}{it}{<-> s * [1.10] pzcmi7t}{}
\DeclareMathAlphabet{\mathpzc}{OT1}{pzc}{m}{it}

\begin{document}

\aboverulesep=0ex
\belowrulesep=0ex

\thispagestyle{empty} \baselineskip=28pt \vskip 5mm
\begin{center} {\LARGE{\bf Multi-Hazard Bayesian Hierarchical Model \\for Damage Prediction}}
	
\end{center}

\baselineskip=12pt \vskip 10mm

\begin{center}\large
    Mary Lai O. Salva\~{n}a\footnote[1]{\baselineskip=10pt Department of Statistics, University of Connecticut, Storrs, CT, USA\\
    E-mail: marylai.salvana@uconn.edu\\}
\end{center}

\baselineskip=17pt \vskip 10mm \centerline{\today} \vskip 15mm

\begin{center}
{\large{\bf Abstract}}
\end{center} 

A fundamental theoretical limitation undermines current disaster risk models: while real-world natural hazards manifest as complex, interconnected phenomena, existing approaches suffer from two critical constraints. First, conventional damage prediction models remain predominantly deterministic, relying on fixed parameters established through expert judgment rather than learned from data. Second, even recent probabilistic frameworks are fundamentally restricted by their underlying assumption of hazard independence, an assumption that directly contradicts the observed reality of cascading and compound disasters. By relying on fixed expert parameters rather than empirical data and treating hazards as independent phenomena, these models dangerously misrepresent the true risk landscape. This work addresses this critical challenge by developing the Multi-Hazard Bayesian Hierarchical Model (MH-BHM), which reconceptualizes the classical risk equation beyond its deterministic origins. The model's core theoretical contribution lies in reformulating a classical risk formula as a fully probabilistic model that naturally accommodates hazard interactions through its hierarchical structure while preserving the interpretability of the traditional hazard-exposure-vulnerability framework. Using tropical cyclone damage data (1952-2020) from the Philippines as a test case, with out-of-sample validation on recent events (2020-2022), the model demonstrates significant empirical advantages: a reduction in damage prediction error by 61\% compared to a single-hazard model, and 80\% compared to a benchmark deterministic model, corresponding to an improvement in damage estimation accuracy of USD 0.8 billion and USD 2 billion, respectively. This improved accuracy enables more effective disaster risk management across multiple domains, from optimized insurance pricing and national resource allocation to local adaptation strategies, fundamentally improving society's capacity to prepare for and respond to disasters.

\par\vfill\noindent
{\bf Some key words:} Bayesian; damages; exposure; multi-hazard; risk; vulnerability.

\clearpage\pagebreak\newpage \pagenumbering{arabic}
\baselineskip=26pt


\section{Introduction} \label{sec:introduction}

Natural hazards represent formidable threats to human populations and infrastructures worldwide, capable of inflicting devastating socio-economic impacts that can reverberate through communities for years or decades. These hazards encompass a broad spectrum of phenomena, including seismic events, volcanic activities, hydrological extremes, and heat-related disasters, each characterized by distinct spatial distribution, temporal evolution, and intensity measurements. Earthquakes are quantified on the moment magnitude scale, while tropical cyclones are categorized by wind speeds and pressure systems. Droughts may persist for months or years, whereas flash floods can transform landscapes in mere hours.

\subsection{Definition of Risk in the Literature}

When these hazards interact with vulnerable, exposed elements, such as populations and infrastructures, the potential for adverse consequences emerges. This potential is called risk \citep{assembly2016report} and is formally defined through the fundamental equation:
\begin{equation}  \label{eqn:risk_formula}
    \textit{Risk} = \textit{Hazard} \times \textit{Exposure} \times \textit{Vulnerability}.
\end{equation}
The equation above expresses how hazards become risks through interaction with vulnerable systems. While a natural hazard is inherently a physical threat, risk materializes only when that hazard threatens vulnerable elements. For instance, an earthquake in an uninhabited region, despite its significant geophysical attributes, generates minimal risk due to the absence of exposed assets or populations. 

The risk equation is extensively used, particularly for economic impact evaluations. \textit{Risk} typically represents expected monetary losses associated with hazard occurrence. \textit{Hazard} characterizes the processes or events that can induce damages. \textit{Exposure} represents the assets that can be subject to the damaging effects of the hazard, such as built infrastructure, economic assets, and population centers. \textit{Vulnerability} models the susceptibility or sensitivity of the exposed assets to the hazard.

The risk formula in (\ref{eqn:risk_formula}) provides a general guideline for building any risk model for any type of hazard. Although the formula shows risk as the product of hazard, exposure, and vulnerability, this formulation essentially illustrates that risk emerges from the interplay of these key components. Various versions of the risk equation appear in the literature. Among them is another widely used risk formula, implemented in CLIMADA, short for CLIMate ADAptation, a popular climate risk modeling platform with an open-source software (\url{https://github.com/CLIMADA-project/climada_python}). In CLIMADA, risk is defined as follows:
\begin{eqnarray} 
    \textit{Risk} & = & \textit{Probability of Occurrence} \times \textit{Damage}, \label{eqn:CLIMADA_risk_formula} \\
    \textit{Damage} & = & \sum_{l = 1}^{L} \textit{Exposure} (\mathbf{s}_l) \cdot f\{\textit{Hazard} (\mathbf{s}_l)\}, \label{eqn:CLIMADA_damage_formula}
\end{eqnarray}
where the symbol ``$\cdot$'' indicates scalar multiplication.
Here $f(\cdot): \mathbb{R} \rightarrow [0, 1]$ is a vulnerability function, also termed fragility function or impact function, that quantifies the probability of damage or the fractional loss of value of the exposed asset given the occurrence of a key stressor, e.g., rain, flood, or wind \citep{emanuel2011global, aznar2019climada, bresch2021climada}. In the CLIMADA risk formula, damage represents the spatially aggregated direct monetary consequences resulting from physical destruction or impairment of assets at exposure sites $\mathbf{s}_1, \mathbf{s}_2, \ldots, \mathbf{s}_L$ during hazard events. These monetary consequences typically reflect the costs required to repair or replace all affected assets. Spatial aggregation is standard practice, as damage reports commonly record a single total value rather than itemized losses.

\subsection{Probabilistic Approaches}

The risk formulas above embody the deterministic approach to risk assessment, where defined inputs produce fixed outputs without incorporating uncertainties or probabilistic variations. This deterministic framework has found widespread application across multiple fields. For example, seismic risk calculations utilize specific values for ground acceleration and structural vulnerability, while agricultural crop insurance models employ fixed parameters for drought intensity and yield loss ratios. Such deterministic formulation, while straightforward, can introduce significant errors that are highly problematic to risk-bearing institutions such as insurers and disaster relief agencies. Probabilistic models address these limitations by incorporating probability distributions and conditional relationships to capture uncertainties in hazards, exposures, and vulnerabilities. For instance, instead of using a single flood height value, probabilistic models employ distributions of possible flood levels to account for meteorological variability \citep{lallemant2021nature, paprotny2021probabilistic, duque2023monte}. Significant contributions extending deterministic modeling to probabilistic frameworks include several noteworthy approaches. \cite{villalta2020spatio} proposed a Bayesian approach to estimate hydro-meteorological risk, measured by a loss variable indicating the number of casualties. They modeled hazard (rainfall) using a spatio-temporal Kriging approach, vulnerability using zero-inflated Poisson or negative binomial distributions, and obtained exposure estimates from population census data. \cite{bodenmann2024accounting} proposed a Bayesian approach to estimate seismic vulnerability functions that accounts for ground motion and vulnerability function parameters uncertainty. \cite{meng2023probabilistic} proposed a machine learning framework called TCP-NGBoost for probabilistic forecasting of tropical cyclone intensity. The model combines LightGBM and NGBoost to generate point predictions and uncertainty estimates for 24-hour intensity changes.

\subsection{Multi-Hazard Approaches}

In various instances in the literature where the aforementioned risk formulas have been employed, only one type of hazard is typically considered. For example, tropical cyclone risk assessments focus on wind speeds while often neglecting concurrent storm surge, rainfall-induced flooding, and landslides \citep{ward2020, eberenz2021regional, do2023multi, baldwin2023vulnerability}; earthquake analyses emphasize ground shaking while overlooking potential earthquake-triggered landslides and tsunamis; flood studies concentrate on inundation while omitting associated riverbank erosion and debris flow; and volcanic hazard assessments consider ash fall while sometimes neglecting concurrent lahars, pyroclastic flows, and toxic gases. While methodologically robust for individual phenomena, this hazard-specific approach may significantly underestimate the total risk by overlooking the compound effects of interacting hazards. To address the limitations of single-hazard approaches, multi-hazard risk assessment frameworks have emerged that explicitly consider the simultaneous or cascading occurrence of multiple hazards. \cite{ming2015quantitative} proposed a multi-hazard risk model to model crop losses in China's Yangtze River Delta region caused by strong wind and flood. \cite{jang2022flood} performed a study using copula, regression, and Monte Carlo methods to analyze the compound influence of rainfall and tide on coastal flood risk. \cite{dietz2022predictions} proposed a hierarchical generalized extreme value probability distribution formulation to model financial damages due to tropical cyclones using the maximum wind speed and the minimum central pressure as the hazards.

\subsection{Vulnerability Function-Based Risk Models}

The vulnerability function is a fundamental component in the aforementioned risk models, providing a mathematically rigorous and interpretable approach to damage estimation. Vulnerability functions model a system's probability of reaching a certain damage level given the intensity of the hazard. These functions can be constructed heuristically, analytically, or empirically \citep{li2013bayesian, lallemant2015statistical, bodenmann2024accounting}. Heuristic methods involve collating opinions from engineers and experts. Judgmental knowledge obtained by this approach is associated with significant uncertainties as experts may have varying quantitative and qualitative damage assessments. Analytical vulnerability functions are developed through detailed structural modeling and computer simulations that use mathematical formulas to predict how buildings and infrastructure respond to hazards. These functions incorporate material properties, structural behavior, and damage metrics to quantify the relationship between hazard intensity and expected damage. Such approaches are computationally demanding as many simulations need to be performed with various configurations in order to thoroughly account for the uncertainties in potential damages \citep{rossetto2005new, besarra2024flood}. Empirical approaches calibrate vulnerability functions using post-disaster damage data. This approach is prevalent in risk assessment and is considered the most realistic.

A widely used vulnerability function is known as the damage power function used in tropical cyclone risk models, proposed in \cite{emanuel2011global}, which has the form:
\begin{eqnarray}
    f(v; v_{\text{thresh}}, v_{\text{half}}) & = & \frac{v^3}{1 + v^3}, \quad v =  \frac{\max (v_o - v_{\text{thresh}}, 0 )}{v_{\text{half}} - v_{\text{thresh}}},
    \label{eqn:emanuel_fragility_function}
\end{eqnarray}
where $v_o$ is the observed wind speed, $v_{\text{thresh}}$ is the wind speed below which no damage is recorded, and $v_{\text{half}}$ is the wind speed at which half of the total value of the asset is lost. The vulnerability function in~(\ref{eqn:emanuel_fragility_function}) implies that damage increases monotonically with wind speed following a sigmoidal function. Another popular class of vulnerability functions is the two-parameter log-normal distribution function, which has been adopted beyond tropical cyclones to model vulnerability due to various hazard types and has the following form:
\begin{eqnarray} \label{eqn:standard_normal_vulnerability_function}
    f(h; \eta, \beta) = \Phi \left(\frac{\ln h - \eta}{\beta} \right),
\end{eqnarray}
where $h$ is the value of the intensity measure of the hazard, $\eta$ is the median parameter, $\beta$ is the standard deviation, and $\Phi(\cdot)$ is the standard normal cumulative distribution function (CDF). Vulnerability functions have also been constructed using generalized linear and additive models \citep{charvet2015multivariate, nguyen2022order}.

In a multi-hazard risk scenario, it is not clear how the vulnerability function in~(\ref{eqn:emanuel_fragility_function}) can be extended to accommodate more than one hazard type. On the other hand, the log-normal vulnerability function model in (\ref{eqn:standard_normal_vulnerability_function}) can easily be extended to accommodate two or more hazards using the multivariate normal CDF ~$\Phi_d(\cdot)$, instead of the univariate~$\Phi(\cdot)$. However, since the multivariate normal CDF has no closed form, one must numerically evaluate $\Phi_d(\cdot)$. The numerical integration can be quite computationally intensive as the dimension $d$ increases \citep{nascimento2022vecchia}. As a multi-hazard extension to the model in~(\ref{eqn:standard_normal_vulnerability_function}), \cite{reed2016multi} and \cite{andriotis2018extended} developed multi-hazard vulnerability functions of the form:
\begin{eqnarray} \label{eqn:logistic_vulnerability_function}
    f(\mathbf{h}; \boldsymbol{\beta}) = \frac{1}{1+e^{g(\mathbf{h}; \boldsymbol{\beta})}},
\end{eqnarray}
where $g(\mathbf{h}; \boldsymbol{\beta}) = \beta_0 + \beta_1 h_1 + \ldots +  \beta_J h_J$. Here $\mathbf{h}$ is the vector of intensity values of different hazards, i.e., $\mathbf{h} = (h_1, \ldots, h_J)^{\top}$, and $\boldsymbol{\beta}$ is a vector of vulnerability parameters, i.e., $\boldsymbol{\beta} = (\beta_0, \beta_1, \ldots, \beta_J)$. Multi-hazard vulnerability functions are now the state-of-the-art and have been used to predict seismic-induced losses \citep{kourehpaz2023toward}, flood-induced losses \citep{nofal2020multi}, tsunami-induced losses \citep{attary2021performance}, and tropical cyclone-induced losses \citep{massarra2020multihazard}. 

Vulnerability function-based risk models have gained widespread adoption across multiple sectors, from insurance catastrophe modeling to government disaster planning, due to their mathematical tractability, clear physical interpretation, and seamless integration with probabilistic hazard assessments. The Federal Emergency Management Agency \citep{FEMA_Hazus_Manuals} Hazus software use vulnerability functions differentiated by building typologies and construction characteristics for earthquakes, floods, and hurricanes. The Florida Public Hurricane Loss Model \citep{FPHLM} utilizes engineering-based vulnerability functions specifically calibrated for hurricane wind and storm surge damage in the Florida. Other prominent examples include the Pacific Earthquake Engineering Research Center \citep{PEER_PBEE_Methodology} framework for seismic risk assessment and the Global Earthquake Model \citep{GEM_Foundation} vulnerability database.

\subsection{Contributions}

Despite its ubiquitous adoption, the classical risk model has remained deterministic in its treatment of hazard, exposure, and vulnerability. The deterministic approach is fundamentally at odds with the inherent uncertainties in each component: hazard intensities vary stochastically, exposure values exhibit strong spatial heterogeneity with high concentrations in urban centers and critical infrastructure hubs, and vulnerability demonstrates significant regional variation, with urban regions typically exhibiting greater structural resilience despite their higher exposure values. Previous approaches have failed to address this intrinsically probabilistic nature, relying instead on deterministic paradigms that can lead to systematic biases, such as damage overestimation when applying country-level vulnerability parameters to well-built urban centers. This study provides the first comprehensive probabilistic treatment of the damage equation in~(\ref{eqn:CLIMADA_damage_formula}) through seven core contributions: (1) reformulation of the damage formula using Bayesian hierarchical modeling, establishing a rigorous statistical approach for uncertainty quantification across all components; (2) incorporation of multi-hazard vulnerability functions to capture complex hazard interactions; (3) development of robust parameter estimation methods integrating empirical data with expert knowledge; (4) implementation of efficient computational algorithms for practical applications; (5) validation of superior damage prediction accuracy compared to traditional deterministic approaches; (6) integration of publicly available hazard and damage datasets into an operational damage prediction model; and (7) first empirical quantification of precipitation's contribution to tropical cyclone damages, validated through historical disaster data, demonstrating both significant multi-hazard effects and the model's capability to capture such interactions. This reformulation provides a complete statistical architecture for treating hazard, exposure, and vulnerability as random variables, fundamentally advancing how uncertainty is quantified in risk assessment.

The remainder of this paper is organized as follows: Section~\ref{sec:proposed} introduces the proposed model; Section~\ref{sec:estimation} presents the procedure for model fitting and prediction, including prior specification, posterior computation, predictive distributions, and algorithmic implementation; Section~\ref{sec:simulation} demonstrates the model's capabilities through simulation studies, comparing its predictive accuracy against established single-hazard approaches across various vulnerability scenarios; Section~\ref{sec:application} validates the proposed model using historical tropical cyclone disaster data, providing empirical evidence of precipitation's contribution to damage alongside wind effects; and Section~\ref{sec:conclusion} concludes with implications for risk assessment practices and directions for future research.



\section{Multi-Hazard Bayesian Hierarchical Model (MH-BHM)} \label{sec:proposed}

Motivated by the disaster risk management community's heavy reliance on vulnerability function-based risk models, the goal is to develop a probabilistic model to predict the magnitude of damage with the multi-hazard vulnerability function as its main engine. This section presents the Multi-Hazard Bayesian Hierarchical Model (MH-BHM), which reformulates the damage formula in (\ref{eqn:CLIMADA_damage_formula}) within a probabilistic framework. The following section reviews the foundations of Bayesian hierarchical models, providing essential background for developing the proposed MH-BHM.

\subsection{Bayesian Hierarchical Model}

A Bayesian hierarchical model (BHM) enables the statistical representation of complex processes through multiple nested stages, each serving a distinct purpose. A standard BHM consists of the following fundamental levels:
\begin{eqnarray*}
    \text{Level 1 - Data Model: } && p(\text{Data}|\text{Process}, \text{Parameters})  \hspace{1cm}  \\
    \text{Level 2 - Process Model: } && p(\text{Process}| \text{Parameters}) \\
    \text{Level 3 - Parameter Model: } && p(\text{Parameters})
\end{eqnarray*}
where $p(A|B)$ indicates the conditional probability distribution of $A$ given $B$ and $p(A)$ denotes the marginal probability distribution of $A$ \citep{wikle2003hierarchical}.
The data model at Level 1 specifies the probability distribution characterizing the relationship between observations and their measurement uncertainties. Level 2 comprises the process model, which defines the underlying mechanisms governing the generation of observed values. Level 3 establishes the parameter model, quantifying uncertainty in model parameters through prior distributions.  

The strength of BHM lies in its ability to decompose complex phenomena through a structured probabilistic framework, with the posterior distribution providing a comprehensive quantification of process and parameter uncertainties. Using Bayes' theorem, the posterior distribution can be expressed as the product of the layers of the BHM, that is,
\begin{eqnarray*}
    p(\text{Process}, \text{Parameters} | \text{Data}) & \propto & \text{Likelihood} \times \text{Prior}, \\
    \text{Likelihood} & = & p(\text{Data}|\text{Process}, \text{Parameters}) \\
    \text{Prior} & = & p(\text{Process}| \text{Parameters}) p(\text{Parameters}),
\end{eqnarray*}
where the symbol $\propto$ means ``proportional to''. The posterior distribution integrates prior knowledge, encoded through prior distributions reflecting initial parameter beliefs, with the likelihood function quantifying the relationship between observations and model parameters.

\subsection{Proposed MH-BHM}

The damage formula in (\ref{eqn:CLIMADA_damage_formula}) can be reformulated as a BHM, where each component contributing to damage calculation is treated as a random variable and modeled at a distinct hierarchical level. This approach allows for comprehensive uncertainty quantification, as uncertainties in each component propagate through the model to influence the final damage prediction. 

Consider a spatial domain $\mathcal{S} \subset \mathbb{R}^2$ where damage-inducing events affect exposed assets. For each event $i = 1,\ldots,N$, the damage at location $s \in \mathcal{S}$ arises when hazard intensity impacts exposed elements according to their vulnerability characteristics. The total damage $D_i$ is obtained by aggregating local damages across the entire spatial domain. To model this aggregation process probabilistically, the classical risk components - hazard, exposure, and vulnerability - are reformulated as stochastic processes. Hazard intensities and asset values are represented by random fields $\mathbf{H}_i(s)$ and $E_i(s)$ respectively, while the vulnerability function $V(\mathbf{H}_i(s))$ maps hazards intensities to damage measures, plus a model error $\epsilon_i$ capturing remaining uncertainties. The components comprising the BHM are defined as follows:
\begin{itemize}
    \item $D_i \in \mathbb{R}^+$ represents the spatially aggregated damage from event $i$, defined as a nonnegative random variable that captures the uncertainty in total damage estimates,
    \item $E_i(s)$ represents the asset value at location $s$, defined as a random field that assigns a random variable to each spatial location, thereby characterizing the spatial correlation and uncertainty of asset distribution,
    \item $\mathbf{H}_i(s) = (H_{i1}(s),\ldots,H_{iM}(s))^\top$ represents the intensities of $M$ hazards at location $s$, defined as a multivariate random field that assigns a vector of correlated random variables to each location, thereby characterizing the spatial correlation and interaction among hazards,
    \item $V(\mathbf{H}_i(s)) \in [0, 1]$ denotes the multi-hazard vulnerability function that maps the joint effect of multiple hazards to a measure of damage at location $s$, and
    \item $\epsilon_i$ is the residual error term capturing both measurement and aggregation uncertainties across events $i = 1,\ldots,N$.
\end{itemize}

\noindent The random fields $E_i(s)$ and $\mathbf{H}_i(s)$ are assumed to be independent, reflecting that asset values do not influence hazard intensities and vice versa. These stochastic components are integrated into a BHM where each level serves a distinct modeling purpose. The data level connects observed damages $D_i$ to the generative processes, while the process level characterizes the spatial behavior of hazards $\mathbf{H}_i(s)$, exposure $E_i(s)$, and vulnerability $V(\mathbf{H}_i(s))$. The parameter level completes the hierarchy by quantifying parameter uncertainties through prior distributions. The formulation at each level is given below.

\subsubsection{Data Model}

For spatially aggregated damage observations $\mathcal{D} = {D_1,\ldots,D_N}$:
\begin{equation}
    D_i = \sum_{s \in \mathcal{S}} E_i(s) \times V(\mathbf{H}_i(s)) + \epsilon_i, \quad \epsilon_i|\boldsymbol{\theta}_\epsilon \sim p(\epsilon_i|\boldsymbol{\theta}_\epsilon).
\end{equation}

\subsubsection{Process Model}

\begin{enumerate}
    \item \textit{Exposure Random Field} \\
    Exposure values are conventionally treated as known, fixed quantities derived from census data \citep{villalta2020spatio, eberenz2021regional, baldwin2023vulnerability}. However, one can treat this component as random and use the following spatial random field model for asset exposure that captures clustering near urban centers:
    \begin{eqnarray}
        E_i(s) & = & \mu_{E_i}(s) + Z_{E_i}(s) \\
        \mu_{E_i}(s) & = & \alpha_0 + \sum_{k=1}^K \alpha_k\exp(-\phi_k|s-c_k|^2) \\
        Z_{E_i}(s) & = & g(W_{E_i}(s)),
    \end{eqnarray}
     where $c_k \in \mathcal{S}$ represents the location of urban center $k$ for $k=1,\ldots,K$, $g(x) = \exp(x)$ is the exponential transformation ensuring non-negative exposure values, $W_{E_i}(s)|\boldsymbol{\theta}_E \sim \mathcal{GP}(0, C_E(s,s'|\boldsymbol{\theta}_E))$ is a Gaussian process such that for any finite collection of locations $s_1,\ldots,s_n \in \mathcal{S}$, $[W_{E_i}(s_1),\ldots,W_{E_i}(s_n)]^\top \sim \mathcal{N}_n(\mathbf{0}, [C_E(s_i,s_j|\boldsymbol{\theta}_E)]_{i,j=1}^n)$, where $\mathcal{N}_{d}(\boldsymbol{\mu}, \boldsymbol{\Sigma})$ denotes the $d$-dimensional multivariate normal distribution with mean vector $\boldsymbol{\mu} \in \mathbb{R}^d$ and covariance matrix $\boldsymbol{\Sigma} \in \mathbb{R}^{d \times d}$. The covariance function $C_E(s,s'|\boldsymbol{\theta}_E)$ can be chosen as the Mat\'{e}rn covariance function, $C(s,s'|\boldsymbol{\theta}_E) = \sigma^2\frac{2^{1-\nu}}{\Gamma(\nu)}\left(\frac{|s-s'|}{\rho}\right)^{\nu} K_{\nu}\left(\frac{|s-s'|}{\rho}\right)$, where $\boldsymbol{\theta}_E = (\sigma^2, \rho, \nu)$, $\sigma^2$ is the variance parameter, $\rho$ is the range parameter, $\nu$ is the smoothness parameter, and $K_{\nu}$ is the modified Bessel function of the second kind.

    \item \textit{Multi-Hazard Random Field} \\
    The multi-hazard random field $\mathbf{H}_i(s)$ is:
    \begin{equation}
        \mathbf{H}_i(s) = \boldsymbol{\mu}_{H_i}(s) + \mathbf{Z}_{H_i}(s),
    \end{equation}
    \sloppy where $\mathbf{Z}_{H_i}(s)|\boldsymbol{\theta}_H \sim \mathcal{MGP}(\mathbf{0}, \mathbf{K}(s,s'|\boldsymbol{\theta}_H))$ is a multivariate Gaussian process such that for any finite collection of locations $s_1,\ldots,s_n \in \mathcal{S}$: $[\mathbf{Z}_{H_i}(s_1)^\top,\ldots,\mathbf{Z}_{H_i}(s_n)^\top]^\top \sim \mathcal{N}_{nM}(\mathbf{0}, [\mathbf{K}(s_i,s_j)]_{i,j=1}^n)$ with cross-covariance function $\mathbf{K}(s,s'|\boldsymbol{\theta}_H) = [K_{jk}(s,s'|\boldsymbol{\theta}_{H_{jk}})]_{j,k=1}^M$ such that $K_{jj}(s,s')$ is a valid covariance function for each hazard $j$, $K_{jk}(s,s') = K_{kj}(s',s)$ for all hazards $j,k$ (symmetry), and the resulting covariance matrix is positive definite. A choice for the cross-covariance function $\mathbf{K}(s,s'|\boldsymbol{\theta}_H) $ is the Mat\'{e}rn cross-covariance function $K_{jk}(s,s'|\boldsymbol{\theta}_{H_{jk}}) = \sigma_{jk}\frac{2^{1-\nu_{jk}}}{\Gamma(\nu_{jk})}\left(\frac{|s-s'|}{\rho_{jk}}\right)^{\nu_{jk}} K_{\nu_{jk}}\left(\frac{|s-s'|}{\rho_{jk}}\right)$ with $\boldsymbol{\theta}_{H_{jk}} = (\sigma_{jk}, \rho_{jk}, \nu_{jk})$ for $j,k=1,\ldots,M$, $\sigma_{jk}$ is the cross-covariance parameter, $\rho_{jk}$ is the spatial range parameter, $\nu_{jk}$ is the smoothness parameter, $K_{\nu_{jk}}$ is the modified Bessel function of the second kind, $\sigma_{jk} = \sigma_{kj}$ and $\rho_{jk} = \rho_{kj}$ and $\nu_{jk} = \nu_{kj}$ (symmetry).

    \item \textit{Multi-Hazard Vulnerability Function} \\
    The multi-hazard vulnerability function models the joint effect of multiple hazards through the following hazard intensity-damage relationship:
    \begin{equation} \label{eqn:multivariate_vulnerability_function}
        V(\mathbf{H}_i(s)) = \frac{1}{1 + \exp(\gamma-\boldsymbol{\beta}^\top\mathbf{H}_i(s))}.
    \end{equation}
    The parameters of the vulnerability function above have the following interpretations: $\gamma$ represents the critical threshold where 50\% of the asset value is lost. This threshold is reached when $\gamma = \boldsymbol{\beta}^\top\mathbf{H}_i(s)$. Higher values of $\gamma$ indicate greater resistance to hazard impacts, while lower values reflect increased susceptibility to damage. $\boldsymbol{\beta}$ captures hazard-specific effects where larger $\beta_j$ indicates stronger impact of hazard $j$, assuming hazards are measured on the same scale and $\beta_j > 0$ implies damage increases with hazard intensity. The multi-hazard vulnerability function above exhibits key properties that ensure realistic damage modeling. The function is bounded between 0 and 1, constraining the proportional damage to physically meaningful values. Under extreme hazard conditions where the hazard intensities significantly exceed the threshold ($\boldsymbol{\beta}^\top\mathbf{H}_i(s) \gg \gamma$), the function approaches 1, representing complete damage. Conversely, when hazard intensities are well below the threshold ($\boldsymbol{\beta}^\top\mathbf{H}_i(s) \ll \gamma$), the function approaches 0, indicating negligible damage.

\end{enumerate}

\subsubsection{Parameter Model} 

The complete set of parameters involved in MH-BHM is defined as follows:

\begin{itemize}[itemsep=0pt,parsep=0pt,topsep=0pt]
    \item Urban center parameters: $\boldsymbol{\alpha} = (\alpha_0,\ldots,\alpha_K) \sim p(\boldsymbol{\alpha})$ and  $\boldsymbol{\phi} = (\phi_1,\ldots,\phi_K) \sim p(\boldsymbol{\phi})$;
    \item Random field parameters: $\boldsymbol{\theta}_E = (\sigma^2_E, \rho_E, \nu_E) \sim p(\boldsymbol{\theta}_E)$ and $\boldsymbol{\theta}_H = (\boldsymbol{\theta}_{H_{jk}})_{j,k=1}^M \sim p(\boldsymbol{\theta}_H)$;
    \item Vulnerability parameters: $\boldsymbol{\beta} \sim p(\boldsymbol{\beta})$ and $\gamma \sim p(\gamma)$; and
    \item Error parameter: $\boldsymbol{\theta}_\epsilon \sim p(\boldsymbol{\theta}_\epsilon)$.
\end{itemize}

\subsubsection{Posterior Distribution} 

The proposed MH-BHM is developed to serve as a tool for predicting damages resulting from concurrent multiple hazards. To utilize the model effectively, one must specify the values of the parameters contained in the vector $\boldsymbol{\theta}$. In real-world applications, practitioners may be uncertain about the appropriate parameter values. However, the availability of relevant data can facilitate the estimation of the parameter values.

Bayesian inference enables data-informed parameter estimation by systematically combining prior knowledge with empirical observations, such as damages from past disasters. Through Bayesian inference, one obtains a probability distribution of plausible parameter values, known as the posterior distribution. This posterior distribution explicitly quantifies parameter uncertainty, revealing the most likely values, their spread, correlations, and the relative likelihood of different parameter configurations. By applying Bayes' theorem, the posterior distribution has the form:
\begin{equation}
    \label{eqn:posterior_distribution}
    p(\boldsymbol{\theta}, \{\mathbf{W}_{E_i}, \mathbf{Z}_{H_i}\}_{i=1}^N|\mathcal{D}) \propto \prod_{i=1}^N p(D_i|\mathbf{E}_i, \mathbf{H}_i, \boldsymbol{\theta}) p(\mathbf{W}_{E_i}|\boldsymbol{\theta}_E) p(\mathbf{Z}_{H_i}|\boldsymbol{\theta}_H) p(\boldsymbol{\theta}),
\end{equation} 
where $\mathbf{W}_{E_i} = \{W_{E_i}(s): s \in \mathcal{S}\}$, $\mathbf{Z}_{H_i} = \{\mathbf{Z}_{H_i}(s): s \in \mathcal{S}\}$, $\mathbf{E}_i = \{E_i(s): s \in \mathcal{S}\}$, $\mathbf{H}_i = \{\mathbf{H}_i(s): s \in \mathcal{S}\}$, for $i = 1, \ldots, N$, and $\boldsymbol{\theta} = \{\boldsymbol{\alpha}, \boldsymbol{\phi}, \boldsymbol{\theta}_E, \boldsymbol{\theta}_H, \boldsymbol{\beta}, \gamma, \boldsymbol{\theta}_\epsilon\}$ is the vector of parameters.

\subsubsection{Posterior Predictive Distribution} 

For a new damage event, the posterior predictive distribution $p(\tilde{D}|\mathcal{D})$ represents the probability of observing new damage $\tilde{D} \in \mathbb{R}^+$ given the observed damage data $\mathcal{D}$. This distribution incorporates uncertainty in parameters, latent processes, and model error. The posterior predictive distribution has the form:
\begin{eqnarray} \label{eqn:posterior_predictive_distribution}
    p(\tilde{D}|\mathcal{D}) & = & \int \int p(\tilde{D}|\boldsymbol{\theta}, \tilde{\mathbf{W}}_E, \tilde{\mathbf{Z}}_H)p(\tilde{\mathbf{W}}_E, \tilde{\mathbf{Z}}_H|\boldsymbol{\theta}, \{\mathbf{W}_{E_i}\}_{i=1}^N, \{\mathbf{Z}_{H_i}\}_{i=1}^N) \nonumber \\ 
    && \times p(\boldsymbol{\theta}, \{\mathbf{W}_{E_i}\}_{i=1}^N, \{\mathbf{Z}_{H_i}\}_{i=1}^N|\mathcal{D})d\tilde{\mathbf{W}}_Ed\tilde{\mathbf{Z}}_Hd\boldsymbol{\theta}d\{\mathbf{W}_{E_i}\}_{i=1}^Nd\{\mathbf{Z}_{H_i}\}_{i=1}^N,
\end{eqnarray}
where $\tilde{\mathbf{W}}_E = \{\tilde{W}_{E}(s): s \in \mathcal{S}\}$ and $\tilde{\mathbf{Z}}_H = \{\tilde{\mathbf{Z}}_{H}(s): s \in \mathcal{S}\}$ are the new latent processes. The computational details on sampling from the posterior predictive distribution in (\ref{eqn:posterior_predictive_distribution}) can be found in the Supplementary Material.

\section{Estimation} \label{sec:estimation}

The posterior distribution of MH-BHM in (\ref{eqn:posterior_distribution}) is intractable, lacking a closed-form solution due to the high-dimensional nature of the parameter space. Thus, the posterior distribution must be obtained using numerical techniques, such as Markov Chain Monte Carlo (MCMC). The MCMC method produces an iterative chain of parameter values $\{\boldsymbol{\theta}_0, \boldsymbol{\theta}_1, \ldots\}$, where each sample depends only on the previous sample.  As the number of iterations increases, the Markov chain approaches its stationary distribution, corresponding to the target posterior distribution.

Among various MCMC algorithms, the Metropolis-Hastings (MH) algorithm is a general framework that constructs a Markov chain with the desired stationary distribution through a proposal-acceptance mechanism.  The algorithm uses a proposal-acceptance mechanism where new parameter values are proposed from a transition kernel and accepted or rejected based on the ratio of the posterior distribution evaluated at the proposed and current parameters. The MH algorithm is particularly well-suited for this model due to its high-dimensional parameter space and complex posterior distribution, which arise from both the nonlinear vulnerability function and the inherent spatial structure. Various proposal distributions can be implemented, such as a random walk with a uniform transition kernel $U(\theta_i - 0.5, \theta_i + 0.5)$ for each component $\theta_i$ of the parameter vector $\boldsymbol{\theta}$.  This symmetric proposal structure offers two key advantages: it simplifies the acceptance ratio computation by canceling out the proposal density terms and maintains an effective balance between parameter space exploration and acceptance rate. For a proposed transition from current parameters included in the chain $\boldsymbol{\theta}$ to proposed parameters $\boldsymbol{\theta}'$, the acceptance probability is $\min\left(1, \frac{p(\mathcal{D}|\boldsymbol{\theta}')p(\boldsymbol{\theta}')}{p(\mathcal{D}|\boldsymbol{\theta})p(\boldsymbol{\theta})}\right)$, where $p(\mathcal{D}|\boldsymbol{\theta})$ denotes the likelihood function and $p(\boldsymbol{\theta})$ represents the prior distribution of the parameters included in the model. 



\section{Simulation Study} \label{sec:simulation}

This section demonstrates the application of the MH-BHM to predict damages from tropical cyclone (TC) events. Synthetic TC events were generated using historical TC records from the Philippines, a region characterized by intense TC activities. Damage values were simulated using the classical damage formula in (\ref{eqn:CLIMADA_damage_formula}) under diverse vulnerability scenarios by systematically varying the parameters of the multi-hazard vulnerability function. The experimental design focused on two objectives: (i) evaluating the estimation procedure's accuracy in recovering the MH-BHM parameters and (ii) quantifying the extent of damage underestimation that occurs when using a single-hazard model to characterize multi-hazard phenomena.

TCs, also known as hurricanes or typhoons, depending on geographical context, are complex meteorological phenomena marked by rotating cloud and thunderstorm systems. They form over warm tropical or subtropical waters, driven by oceanic heat transfer and initiated by pre-existing weather disturbances \citep{emanuel2003tropical}. TCs are categorized based on the Saffir-Simpson Hurricane Wind Scale, which classifies storms according to their maximum sustained wind speeds. The scale distinguishes between low-intensity TCs (categories 1 and 2) and high-intensity TCs (categories 3, 4, and 5), with category 5 TCs designated as supertyphoons \citep{bourdin2022intercomparison}.

TCs represent one of the most destructive natural disaster types globally and are comprised of four primary hazards: destructive winds, storm surges, heavy rainfall-induced flooding, and landslides. \citep{ do2023multi, vogt2024modeling}. Despite the recognized importance of assessing TC impacts on communities, infrastructure, and natural resources, existing TC risk models in the literature remain limited, naive, and overly simplistic. The proposed MH-BHM represents a significant advancement over these existing approaches, offering a more comprehensive and robust framework for TC risk assessment.

\subsection{Synthetic Data Generation}

The procedure for simulating damage values from synthetic TC events is outlined below.
\begin{enumerate}

\begin{figure}[t]
            \begin{subfigure}[b]{0.48\textwidth}
                \centering
                \includegraphics[width=\linewidth]{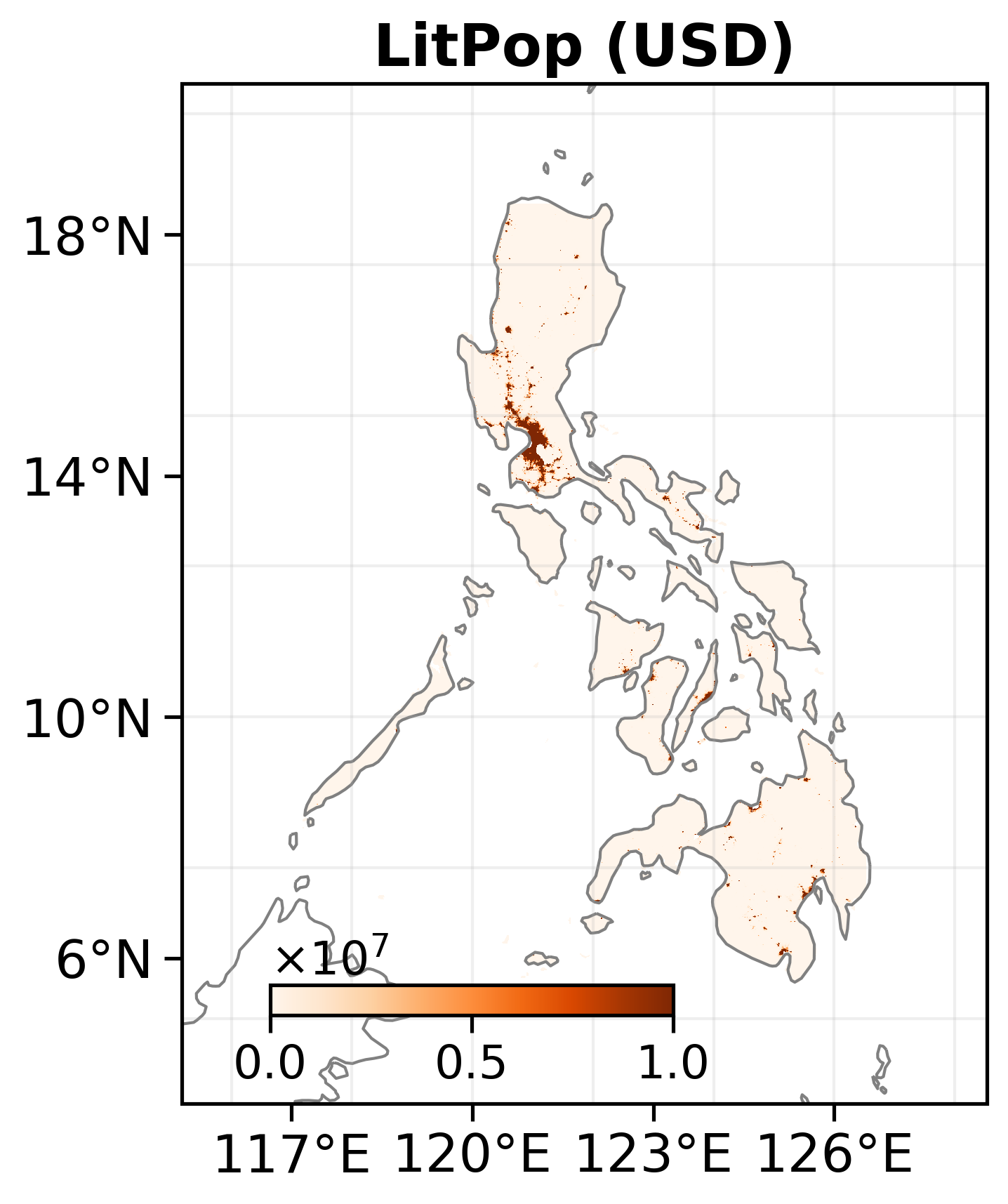}
                \caption{}
                \label{fig:fig-exposure-ph}
            \end{subfigure}%
            \hspace{\fill}
            \begin{subfigure}[b]{0.48\textwidth}
                \centering
                \includegraphics[width=\linewidth]{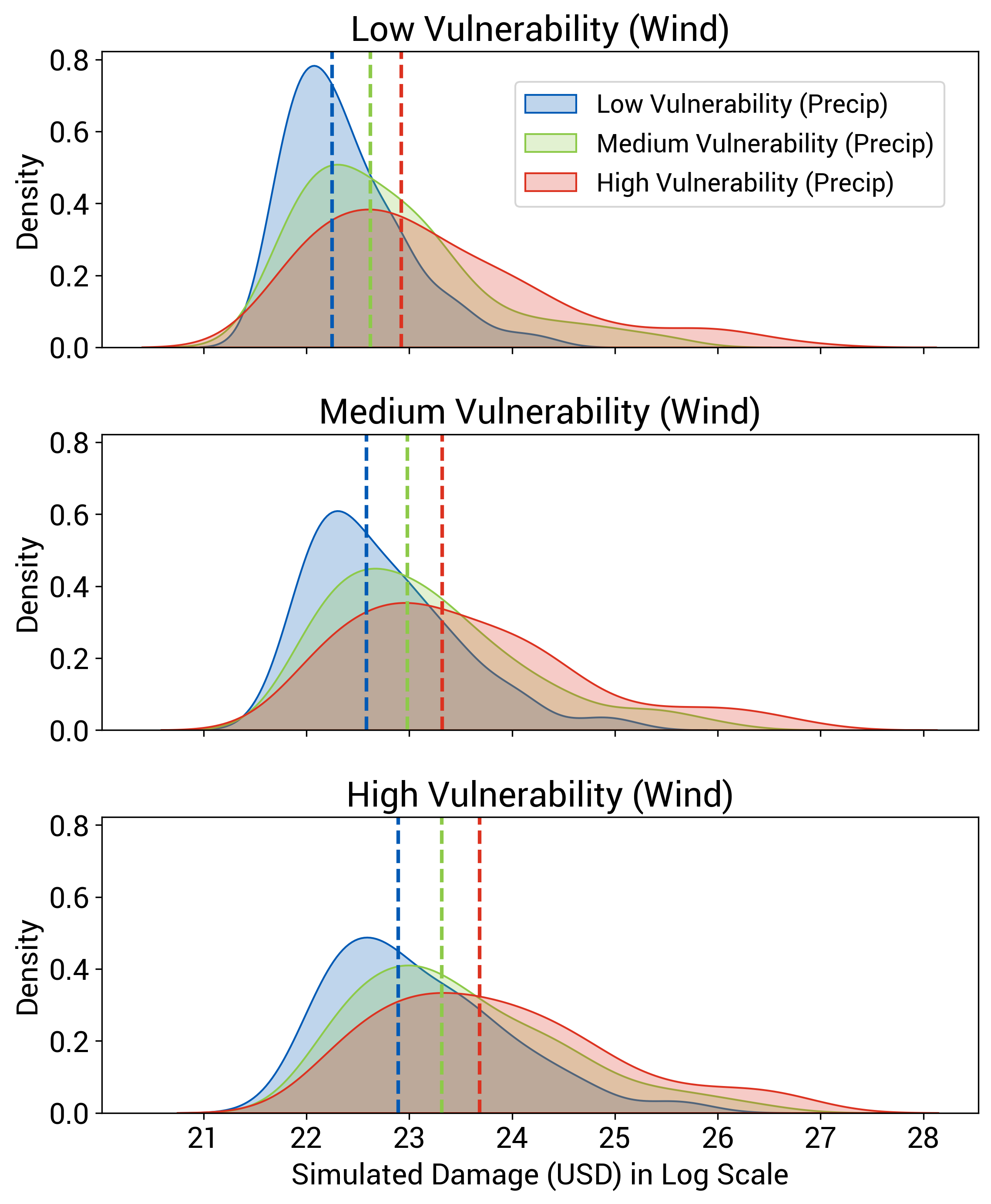}
                \caption{}
                \label{fig:fig-damages-simulated-histogram}
            \end{subfigure}
            \caption{(a) Gridded exposure values (2014 prices). (b) Empirical distribution of the simulated damages in logarithmic scale, under various vulnerability scenarios, with vertical lines indicating the median damage value.} 
        \end{figure}
        
    \item (Exposure) Obtain the exposure data for the Philippines. \\
    The Philippines' physical gridded asset exposure values were obtained from the LitPop module in CLIMADA. The dataset comes at a high resolution of 30-arc-s ($\sim$1 kilometer). Figure~\ref{fig:fig-exposure-ph} depicts the heatmap of asset values across the Philippines. These values were derived using the LitPop method \citep{eberenz2020asset}. The LitPop approach allocates asset values to each grid point based on the product of nightlight intensity (Lit) and population data (Pop) at each grid point. High asset exposure values concentrate around Manila, the country's capital city. 

    \item (Hazards) Generate wind and precipitation fields.
    \begin{enumerate}
        \item Generate an ensemble of synthetic TC tracks from historical TC tracks using the Monte Carlo-based TC emulator of \cite{nederhoff2024accounting}. Figure~\ref{fig:tc-tracks} depicts the historical and synthetic tracks of landfalling TCs in the Philippines.

        \begin{figure}
            \begin{subfigure}[b]{0.5\textwidth}
            \centering
                \includegraphics[width=0.8\linewidth]{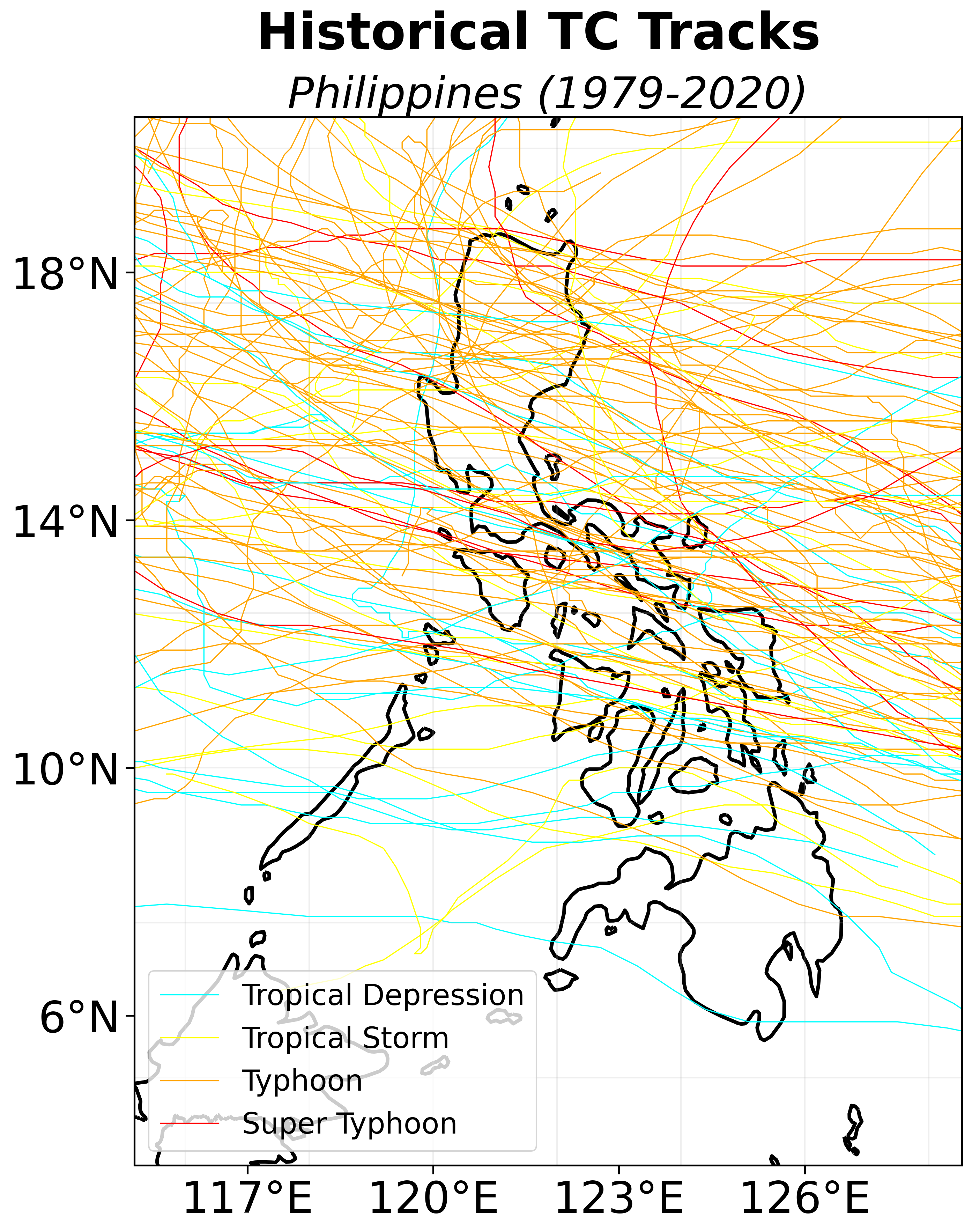}
                \caption{}
                \label{fig:fig-historical-tc-tracks-ph}
            \end{subfigure}%
            \hspace{\fill}
            \begin{subfigure}[b]{0.5\textwidth}
            \centering
                \includegraphics[width=0.8\linewidth]{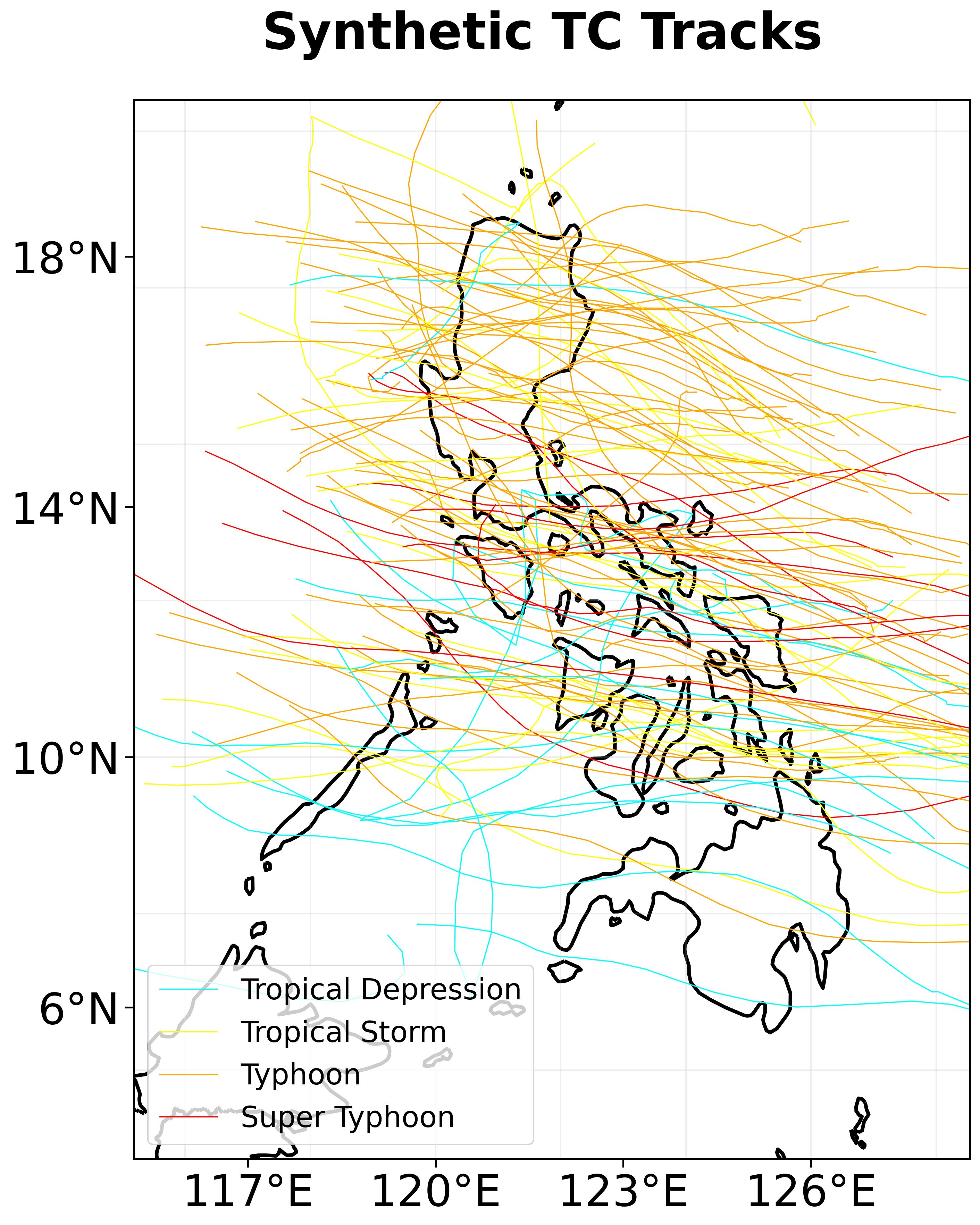}
                \caption{}
                \label{fig:fig-ph-synthetic-tc-tracks}
            \end{subfigure}
            \caption{(a) Historical and (b) synthetic TC tracks of 113 landfalling TCs in the Philippines.} \label{fig:tc-tracks}
        \end{figure}

        \item For each synthetic TC track, generate the hydrometeorological variables: wind and precipitation. These variables represent the hazards destroying exposed assets, with higher values leading to increased potential damage. Wind and precipitation fields were generated using essential TC track parameters, including storm position, central pressure, environmental pressure, radius of maximum winds, and translation speed. These parameters provide inputs for creating a comprehensive spatial representation of wind speeds and precipitation surrounding the TC's center. Detailed methodological approaches are elaborated in \cite{nederhoff2024accounting}. Figure~\ref{fig:tc_tracks_training} visualizes the wind and precipitation fields of the two most destructive TCs in the Philippines. 
        \item Bilinearly interpolate the wind and precipitation fields to match the higher spatial resolution of the exposure data.
    \end{enumerate}

    \item (Vulnerability) Evaluate the multi-hazard vulnerability function at every location $s$:
    \begin{eqnarray} \label{eqn:multivariate_vulnerability_function}
    V(\mathbf{H}(s)) = \frac{1}{1 + \exp(\gamma-\beta_{wind}H_{wind}(s)-\beta_{precip}H_{precip}(s))}.
\end{eqnarray}

    \begin{figure}[t]
    \centering
        \includegraphics[width=0.65\linewidth]{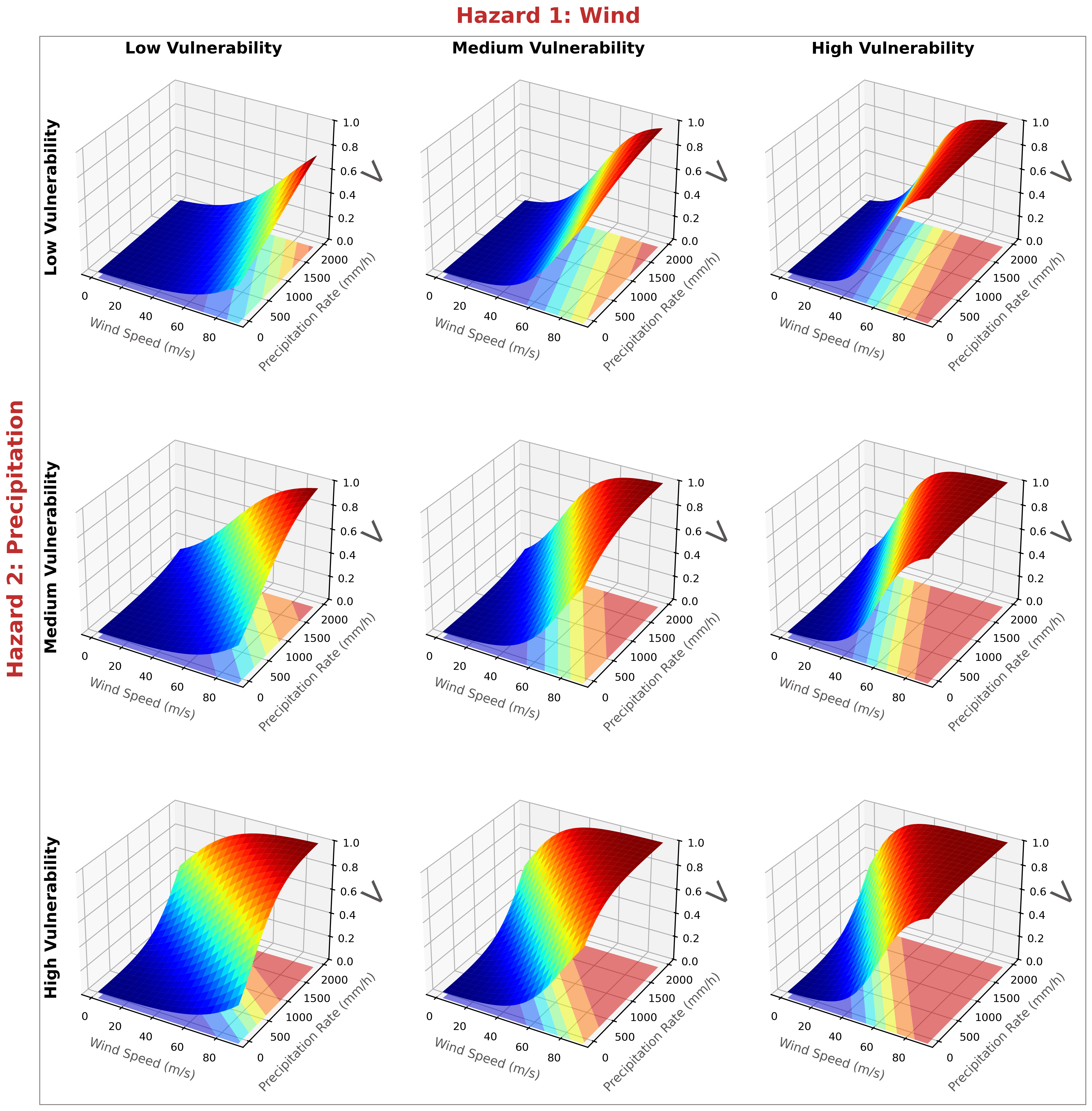}
        \caption{3D surface plots of the multi-hazard vulnerability function under different vulnerability scenarios. The vertical axis represents the proportion of asset value destroyed by the hazards.}
        \label{fig:surface_plot_vulnerability_functions}
\end{figure}
    
    Figure~\ref{fig:surface_plot_vulnerability_functions} visualizes the multi-hazard vulnerability function under different combinations of parameter values, representing the high, medium, and low vulnerability scenarios. The $z$-axis represents the proportion of asset value destroyed by the hazards. The plot shows that increasing vulnerability to a single hazard results in a higher fraction of asset value lost. Furthermore, increased vulnerability to both hazards can lead to greater asset destruction.
    
    \item (Damage) Compute the theoretical damage using the CLIMADA damage formula in (\ref{eqn:CLIMADA_damage_formula}), with multiplicative log-normal error $\exp(\epsilon)$, i.e., $\epsilon \sim \mathcal{N}(0, 3)$. Figure~\ref{fig:fig-damages-simulated-histogram} illustrates the empirical distributions of the simulated damage values in logarithmic scale across vulnerability scenarios. A rightward shift of both the distributions and the medians (vertical lines) can be seen in the figure, demonstrating that damages increase with increasing vulnerability to both wind and precipitation. 
    
\end{enumerate}

\subsection{Tropical Cyclone MH-BHM (TC MH-BHM)}

For the simulation study, the MH-BHM is specialized for TCs and termed the TC MH-BHM. The TC MH-BHM adopts conventional practices by treating hazard and exposure components as known while concentrating on the uncertainty of the vulnerability component. By maintaining deterministic inputs for hazards and exposure, the model aligns with established TC risk modeling practices \citep{eberenz2020asset, baldwin2023vulnerability}. The TC MH-BHM is as follows:
{\allowdisplaybreaks
\begin{alignat*}{2} \label{eqn:TC_MH_BHM}
\text{\footnotesize (Data Model)} & \quad \quad \log(D_i) | E(s), \mathbf{H}_i(s), \boldsymbol{\beta}, \gamma, \sigma^2 \stackrel{ind}{\sim} \mathcal{N}_1\left(\log\left(\sum_{s \in \mathcal{S}} E(s) \times V(\mathbf{H}_i(s))\right), \sigma^2\right), \nonumber \\
\text{\footnotesize (Vulnerability Function)} & \quad \quad V(\mathbf{H}_i(s)) = \frac{1}{1 + \exp(\gamma-\beta_{wind}H_{i,wind}(s)-\beta_{precip}H_{i,precip}(s))}, \nonumber \\
\text{\footnotesize (Exposure)} & \quad \quad E(s) = \text{known exposure at location } s \text{ (constant across TC events)}, \nonumber \\
\text{\footnotesize (Hazard)} & \quad \quad \mathbf{H}_i(s) = \text{known hazard intensities for TC event } i \text{ at } s, & \hspace{4cm} \nonumber \\
\text{\footnotesize (Parameter Model)} & \quad \quad \beta_{wind} \sim \text{Gam}(5,1), \; \beta_{precip} \sim \text{Gam}(5,1), \; \gamma \sim \text{U}(5,15), \; \sigma^2 \sim \text{Gam}(2,0.5), \nonumber \\
\text{\footnotesize (Posterior)} & \quad \quad p(\beta_{wind}, \beta_{precip}, \gamma, \sigma^2|\mathcal{D}) \propto p(\mathcal{D}|\beta_{wind}, \beta_{precip}, \gamma, \sigma^2) p(\beta_{wind})p(\beta_{precip})p(\gamma)p(\sigma^2), \nonumber \\
\text{\footnotesize (Posterior Predictive)} & \quad \quad p(\tilde{D}|\mathcal{D}) = \int p(\tilde{D}|E(s), \tilde{\mathbf{H}}(s), \beta_{wind}, \beta_{precip}, \gamma, \sigma^2)\nonumber \\
& \quad \quad \times p(\beta_{wind}, \beta_{precip}, \gamma, \sigma^2|\mathcal{D})d\beta_{wind}d\beta_{precip}d\gamma d\sigma^2, 
\end{alignat*}
}
where $i = 1, \ldots, N$, $N$ is the number of TC events. Given that damage represents nonnegative economic losses, a log-normal distribution is adopted, denoted by $\mathcal{LN}$. Moreover, the exposure field $E(s)$ is assumed constant across events, while hazard fields $\mathbf{H}_i(s)$ vary by event due to unique tropical cyclone characteristics and are obtained from TC emulators. 

The MCMC simulations were performed in \texttt{R} with three parallel chains using different initial values to ensure convergence. Each chain was run for 5,000 iterations, with the first 2,000 iterations discarded as burn-in. Posterior means were estimated using sample means from the post-burn-in period. Convergence was assessed through visual inspection of trace plots for chain mixing and confirmation that the Gelman-Rubin diagnostic statistic approached 1.

\subsection{Results}

\subsubsection{Accuracy of Parameter Estimates}

\begin{table}[t]
    \centering
    \caption{Summary Statistics of MCMC Parameter Estimates}
    \label{tab:mcmc_summary}
    \begin{tabular}{lccccccc}
        \hline
        Parameter & True Value & Mean & Median & Std Dev & 2.5\% & 97.5\% & $\hat{R}$ \\
        \hline
        $\gamma$ & 6.0 & 6.069 & 6.0508 & 0.4282 & 5.2907 & 6.9523 & 1.0015 \\
        $\beta_{\text{wind}}$ & 7.0 & 7.4744 & 7.4206 & 1.7439 & 4.2504 & 11.101 & 1.0017 \\
        $\beta_{\text{precip}}$ & 6.0 & 5.6041 & 5.5594 & 1.1117 & 3.5982 & 8.0008 & 1.0000 \\
        $\sigma^2$ & 3.0 & 2.8927 & 2.8562 & 0.4061 & 2.2132 & 3.7903 & 1.0000 \\
        \hline
    \end{tabular}
    \begin{tablenotes}
      \small
      \item Summary statistics and convergence diagnostics for MCMC parameter estimation in medium wind-high precipitation vulnerability scenario, showing posterior estimates, credible intervals, and Gelman-Rubin statistic ($\hat{R}$) for each parameter.
    \end{tablenotes}
\end{table}

\begin{figure}[t!]
	\centering
	\includegraphics[width=0.9\textwidth]{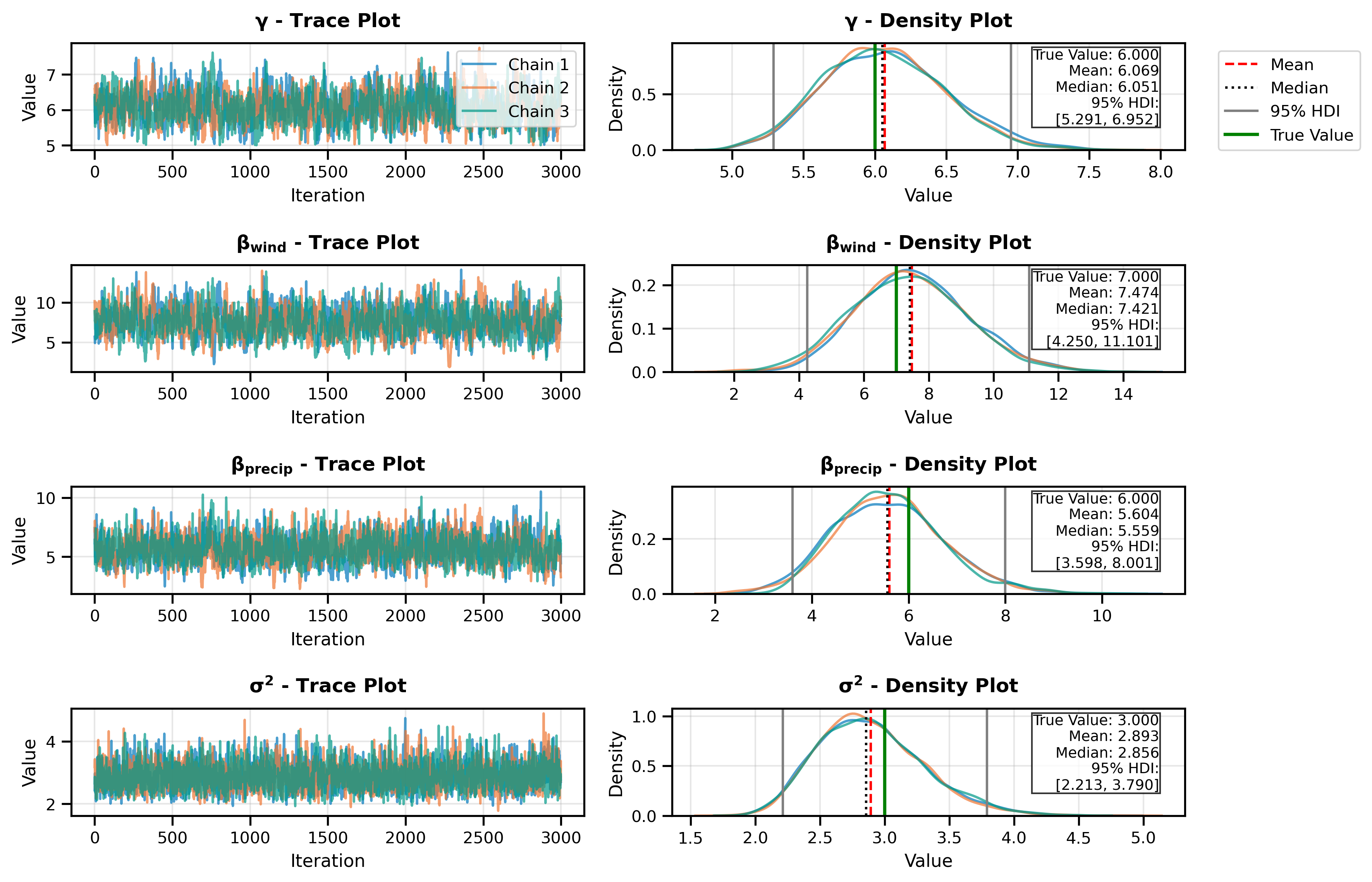}
     \caption{MCMC Parameter Estimation Results}
     \label{fig:mcmc_results_synthetic}
\end{figure}

Figure~\ref{fig:mcmc_results_synthetic} shows key model parameters' convergence diagnostics and posterior distributions. Table~\ref{tab:mcmc_summary} supplements these findings with detailed summary statistics. The trace plots (left panels) demonstrate good mixing and convergence across all chains, which is confirmed by Gelman-Rubin statistics~($\hat{R}$) near 1 for all parameters ($\hat{R} \leq 1.0017$). The complete results for all nine scenarios can be found in the Supplementary Material.

\subsubsection{Damage Prediction Performance}

To evaluate the TC MH-BHM's predictive performance, an out-of-sample validation was conducted using synthetic TC tracks generated for the period 2020-2022. These synthetic tracks were simulated following historical TC patterns during the forecast horizon. For each track, damage values were simulated to match the size of the posterior predictive distribution from the fitted models.
The out-of-sample risk assessment employs multiple complementary metrics. The Value-at-Risk (VaR) at confidence level $\alpha$ represents the damage threshold that will not be exceeded with probability $\alpha$, calculated as,
$\text{VaR}_{\alpha} = F^{-1}(\alpha), $
where $F^{-1}$ is the inverse cumulative distribution function of damages. The Tail Value-at-Risk (TVaR), also known as Expected Shortfall (ES), measures the expected loss beyond VaR, computed as,
$ \text{TVaR}_{\alpha} = E[X|X > \text{VaR}_{\alpha}], $
where $X$ represents damages. The exceedance probability curves complement these metrics by visualizing the probability of exceeding any given damage threshold, calculated as,
$ P(X > x) = 1 - F(x), $
where $F$ is the empirical cumulative distribution function of damages. These metrics collectively provide a comprehensive assessment of model performance across different aspects of the damage distribution, with particular emphasis on capturing extreme events in the distribution tails. To assess distributional similarity between predicted and true damages, the Wasserstein distance is utilized, defined as,
$ W(P,Q) = \inf_{\gamma \in \Gamma(P,Q)} \int ||x-y|| d\gamma(x,y), $
where $P$ and $Q$ are the probability distributions of predicted and true damages, respectively, and $\Gamma(P,Q)$ represents the set of all joint distributions with marginals $P$ and $Q$. 

\begin{figure}[t]
    \centering
    \begin{subfigure}[b]{0.49\textwidth}
        \centering
        \includegraphics[width=0.9\linewidth]{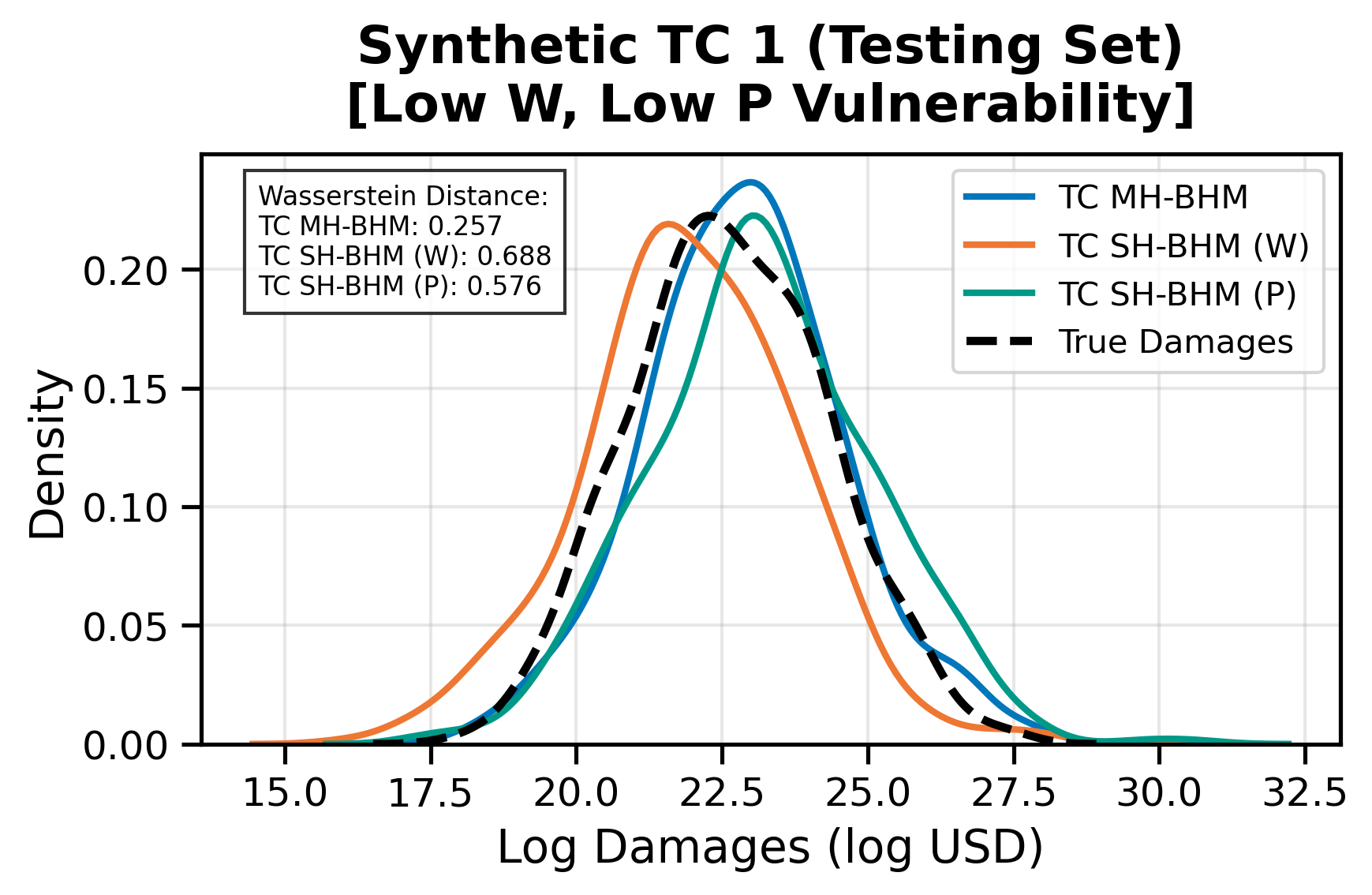}
        \includegraphics[width=0.9\linewidth]{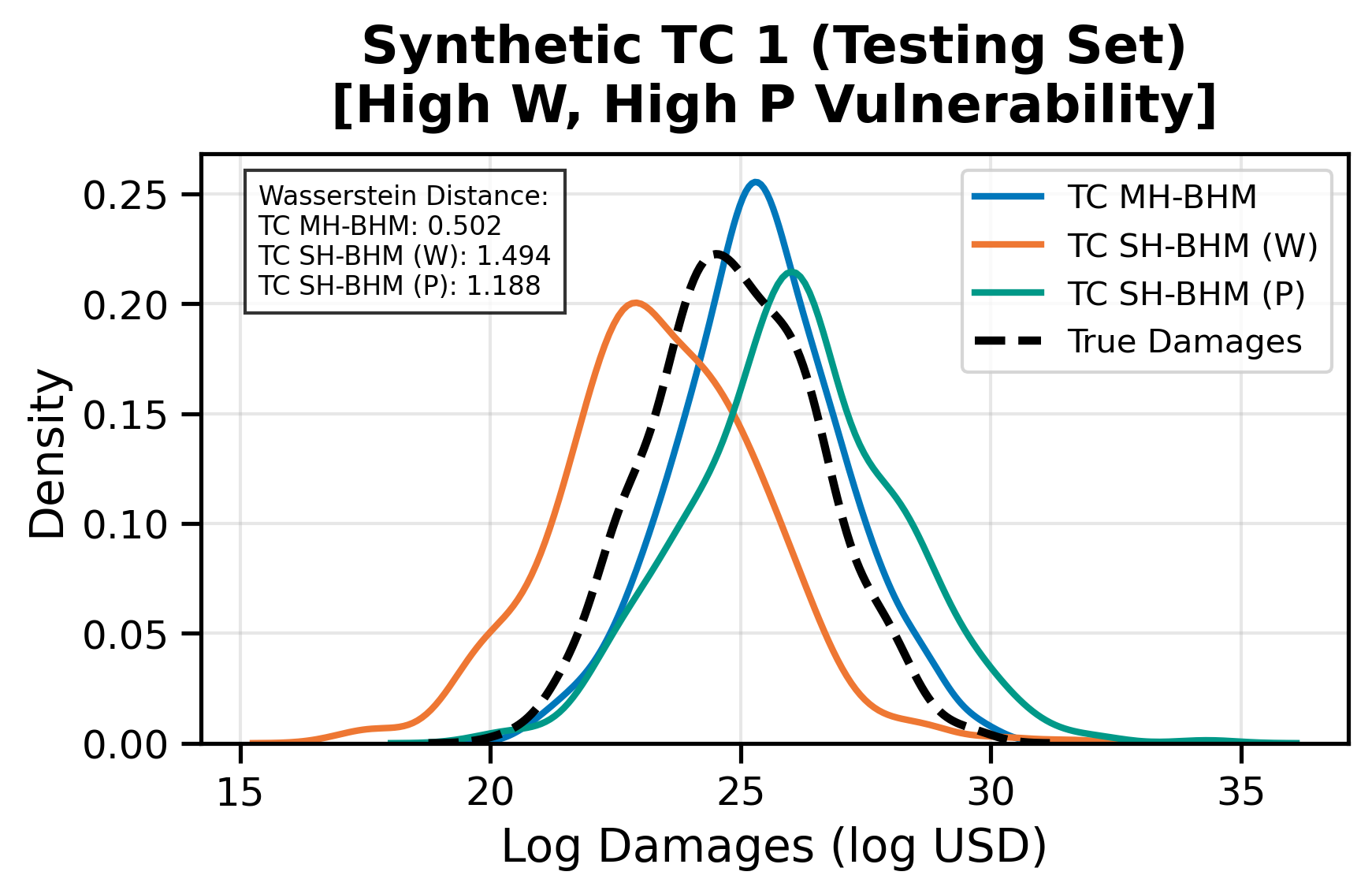}
        \caption{}
    \end{subfigure}%
    \hspace{\fill}
    \begin{subfigure}[b]{0.49\textwidth}
        \centering
        \includegraphics[width=0.9\linewidth]{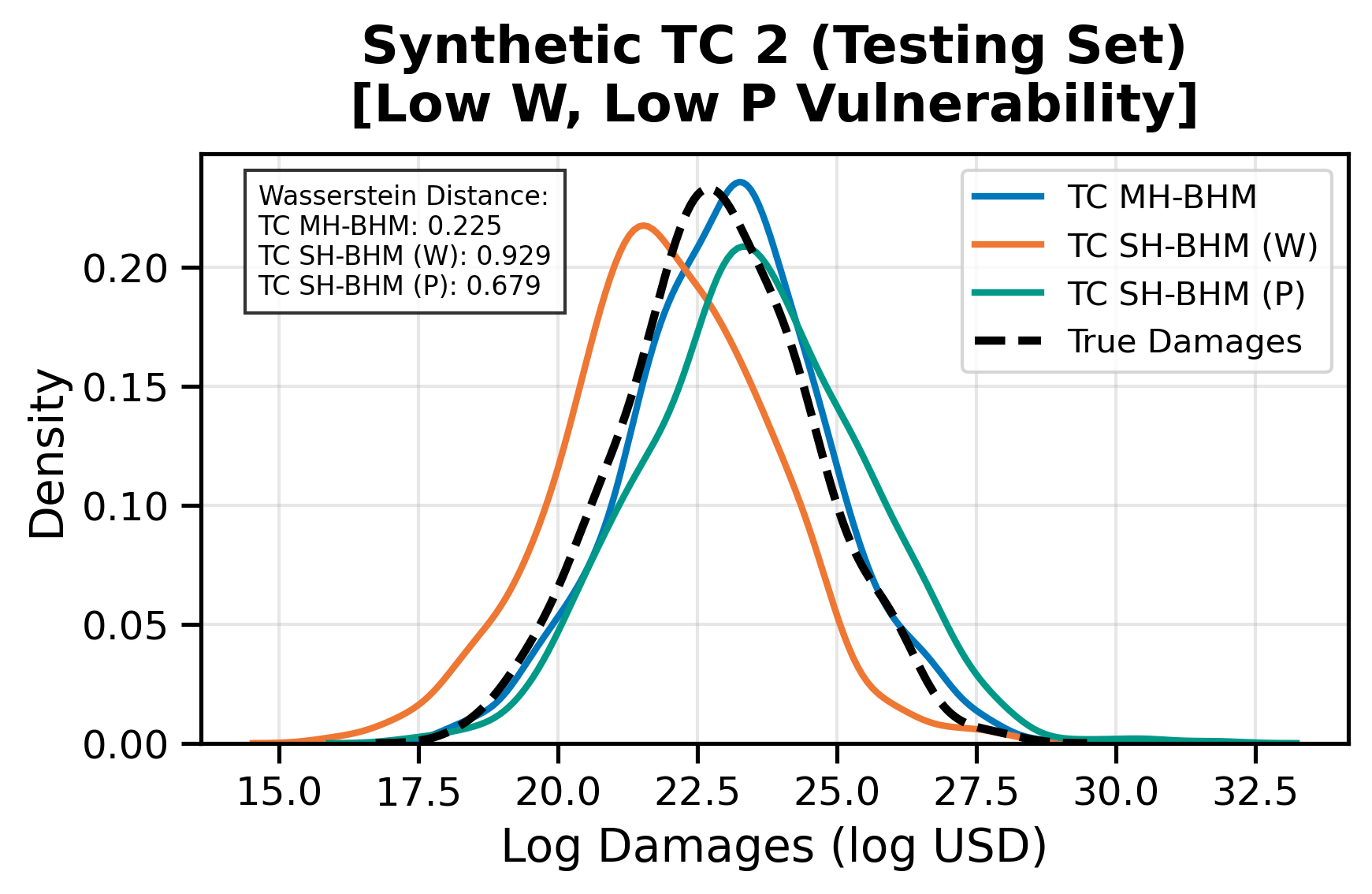}
        \includegraphics[width=0.9\linewidth]{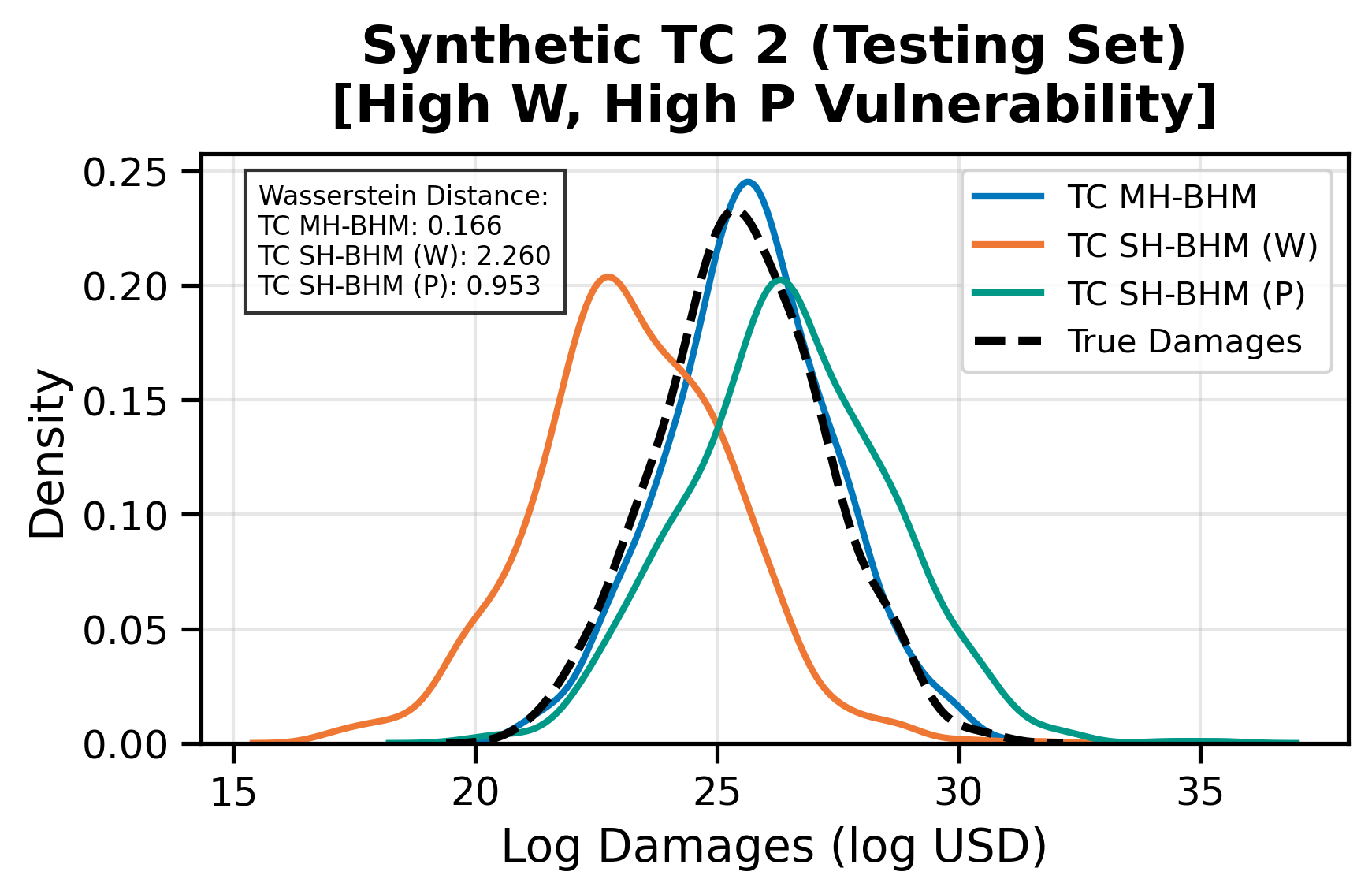}
        \caption{}
    \end{subfigure}
    \caption{Out-of-sample posterior predictive distributions comparing damage predictions from multi-hazard (blue), single-hazard wind speed-only (orange), and single-hazard precipitation-only (green) models against true damage distribution (black) under low and high vulnerability scenarios for two synthetic TC events in the testing set.} 
    \label{fig:fig-posterior-predictive-densities}
\end{figure}

Figure~\ref{fig:fig-posterior-predictive-densities} presents the empirical distributions of damage predictions from the three models TC MH-BHM (multi-hazard), TC SH-BHM (W) (wind-based), and TC SH-BHM (P) (precipitation-based) compared against the true damage empirical distribution for two synthetic TCs in the testing set under varying vulnerability scenarios. From the figure, TC MH-BHM emerges as the best model because it consistently produces damage distributions that most closely align with the true damage patterns, as evidenced by the lowest Wasserstein distances across all scenarios. Moreover, the plots reveal a notable pattern: while all models perform reasonably well under low vulnerability conditions, the discrepancies between model predictions become increasingly pronounced as vulnerability levels escalate. This divergence is particularly evident in the high vulnerability scenarios, where TC SH-BHM (W) and TC SH-BHM (P) show substantial deviations from the true damage distribution. The widening gap between single-hazard and multi-hazard model performances under heightened vulnerability conditions strongly indicates that ignoring the compounding effects of multiple hazards can lead to significant underestimation or overestimation of damages, particularly in highly vulnerable regions.

\begin{figure}[t]
    \begin{subfigure}[b]{0.48\textwidth}
                \centering
                \includegraphics[width=\linewidth]{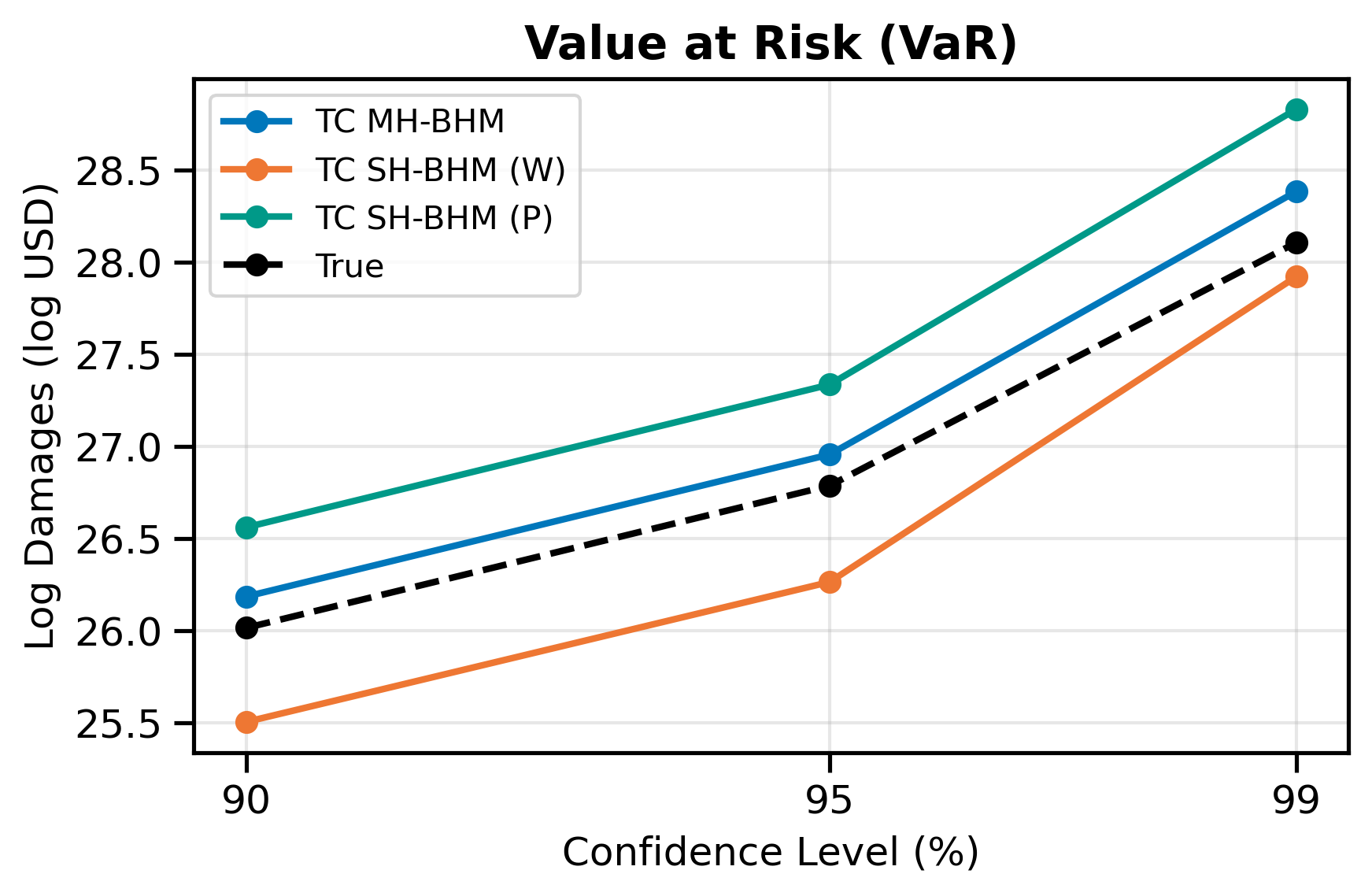}
                \caption{}
                \label{fig:var}
            \end{subfigure}%
            \hspace{\fill}
            \begin{subfigure}[b]{0.48\textwidth}
                \centering
                \includegraphics[width=\linewidth]{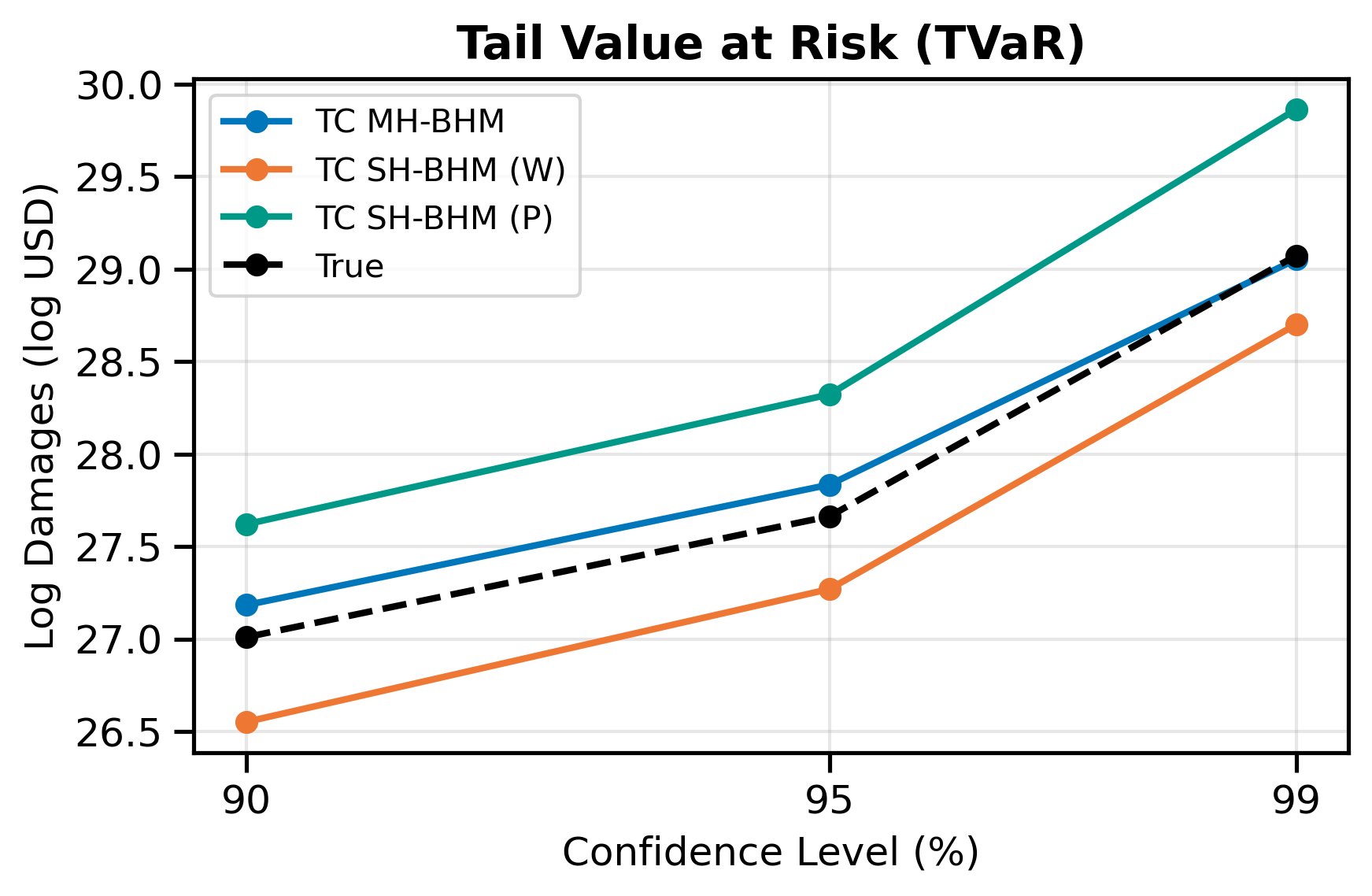}
                \caption{}
                \label{fig:tvar}
            \end{subfigure}
            \begin{subfigure}[b]{0.48\textwidth}
                \centering
                \includegraphics[width=\linewidth]{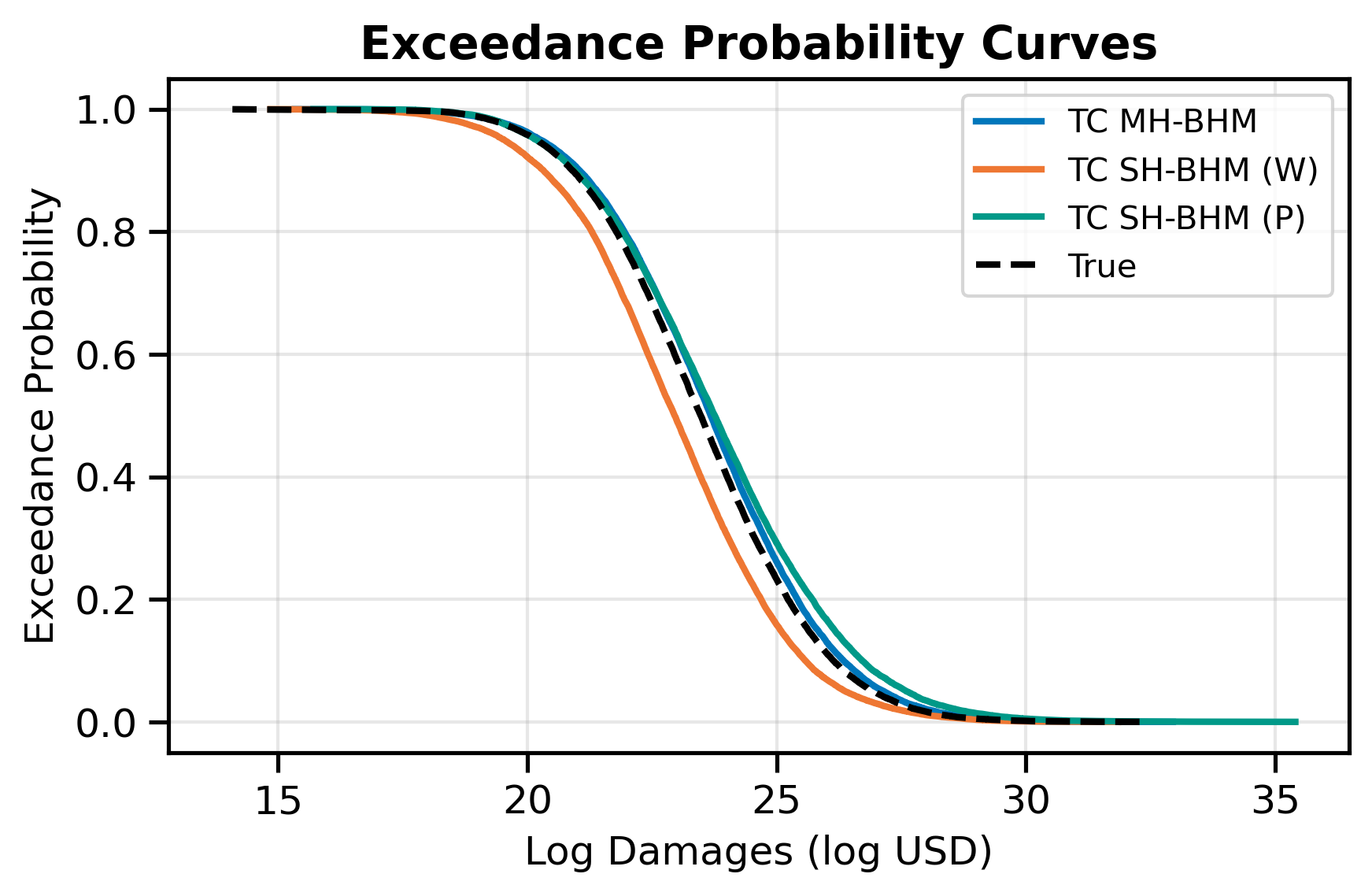}
                \caption{}
                \label{fig:exceedance}
            \end{subfigure}%
            \hspace{\fill}
            \begin{subfigure}[b]{0.48\textwidth}
                \centering
                \includegraphics[width=\linewidth]{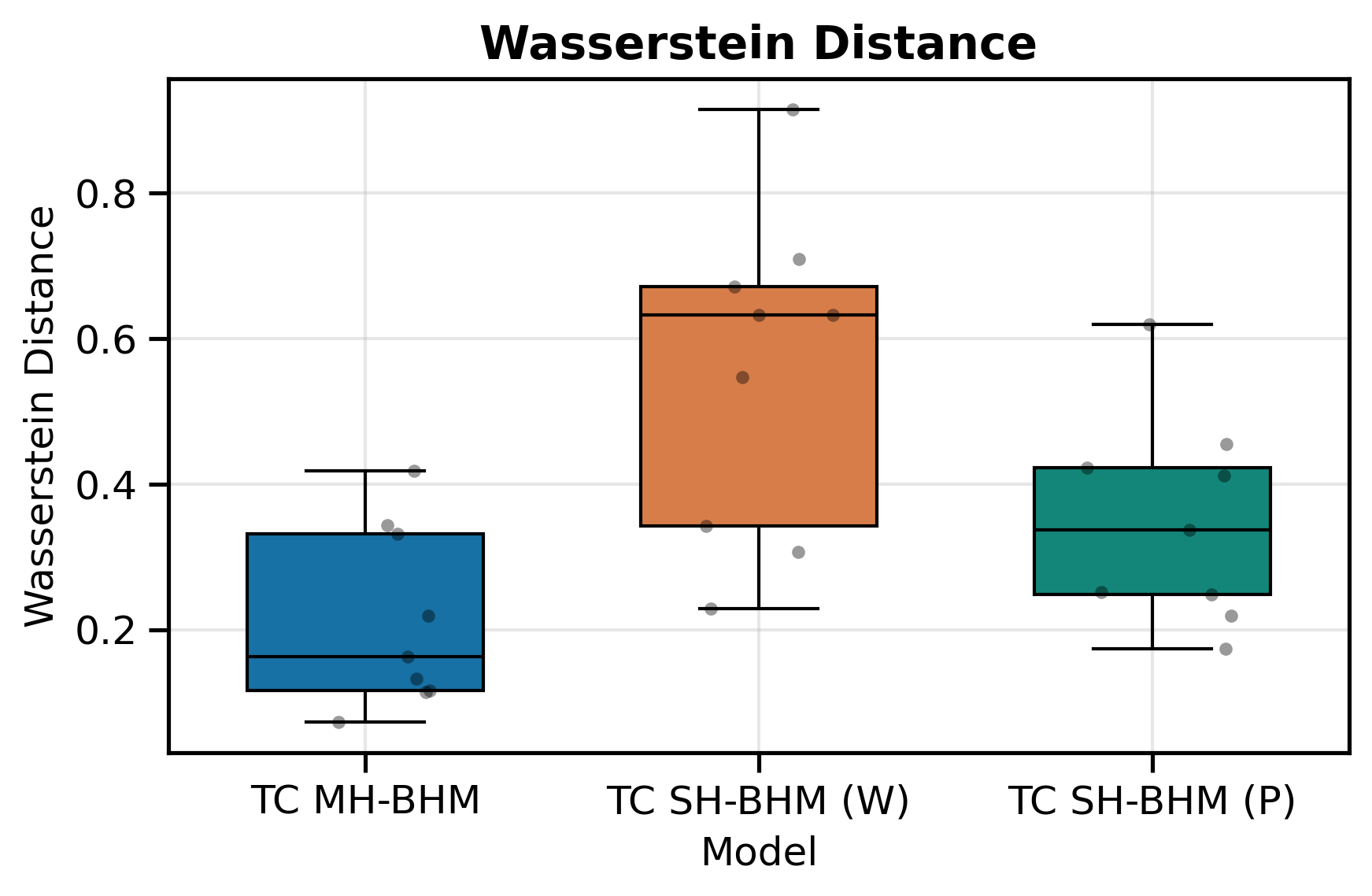}
                \caption{}
                \label{fig:wasserstein}
            \end{subfigure}
    \caption{Risk assessment metrics across competing models.}
    \label{fig:risk_metrics}
\end{figure}

Figure~\ref{fig:risk_metrics} presents various model results for each risk metric. Based on the charts, the TC MH-BHM appears to be the best-performing model among the options presented. Figure~\ref{fig:var} shows the VaR estimates across confidence levels (90\%, 95\%, 99\%) for the three models against VaR estimates of the true model. VaR represents the minimum potential loss at a given confidence level. For example, a 95\% VaR indicates the loss threshold that would not be exceeded 95\% of the time. The figure shows that TC MH-BHM is the best model as it provides a balanced and slightly conservative estimation of VaR across all confidence levels. While TC SH-BHM (W) consistently underestimates the true VaR, potentially leading to inadequate risk preparation, and TC SH-BHM (P) shows excessive overestimation across all levels, showing excessive conservatism that might lead to over-allocation of resources for risk management, TC MH-BHM maintains a moderate and consistent overestimation that offers appropriate risk protection without being overly conservative. Moreover, at the 90\% confidence level, the difference between TC MH-BHM and true model VaR values is $\approx$ \$37 billion, while TC SH-BHM (W) shows a gap of $\approx$ $-$\$80 billion and TC SH-BHM (P) exhibits a difference of $\approx$ \$143 billion. These TC MH-BHM results translate to potential cost savings of \$106 billion compared to TC SH-BHM (P) and improved risk coverage of \$80 billion compared to TC SH-BHM (W). Similar results can be found for the TVaR metric as shown in Figure~\ref{fig:tvar}.

The exceedance probability curves in Figure~\ref{fig:exceedance} reveal that TC MH-BHM and TC SH-BHM (P) best approximate the true damage distribution, with TC MH-BHM showing exceptional alignment across damage levels, particularly in the tail region. The Wasserstein distances in Figure~\ref{fig:wasserstein} further confirms TC MH-BHM's superior performance, consistently demonstrating the lowest distributional deviation from true damage values.


\section{Application} \label{sec:application}

Owing to its location in the western Pacific Ocean, the Philippines gets hit by more TCs than any other country. 
Of all TCs that form globally each year, about 25\% of them, which on average is 20 TCs, develop in the West Pacific (WP) basin, with approximately 8 to 9 of them making landfall in the Philippines \citep{basconcillo2021recent, santos20212020, PAGASA}. Destructive TCs, on average, cost the Philippines 355 million USD annually and claim about 102 lives per event \citep{yonson2018measurement}. Predicting the impact of these TCs is a critical component of the National Resilience Council's anticipatory technologies, designed to build the Philippines' resilience against multiple hazards and cascading disasters \citep{Tangonan2024AnticipatoryCF}.

\begin{figure}
        \begin{subfigure}[b]{0.48\textwidth}
        \centering
                \includegraphics[scale=0.23]{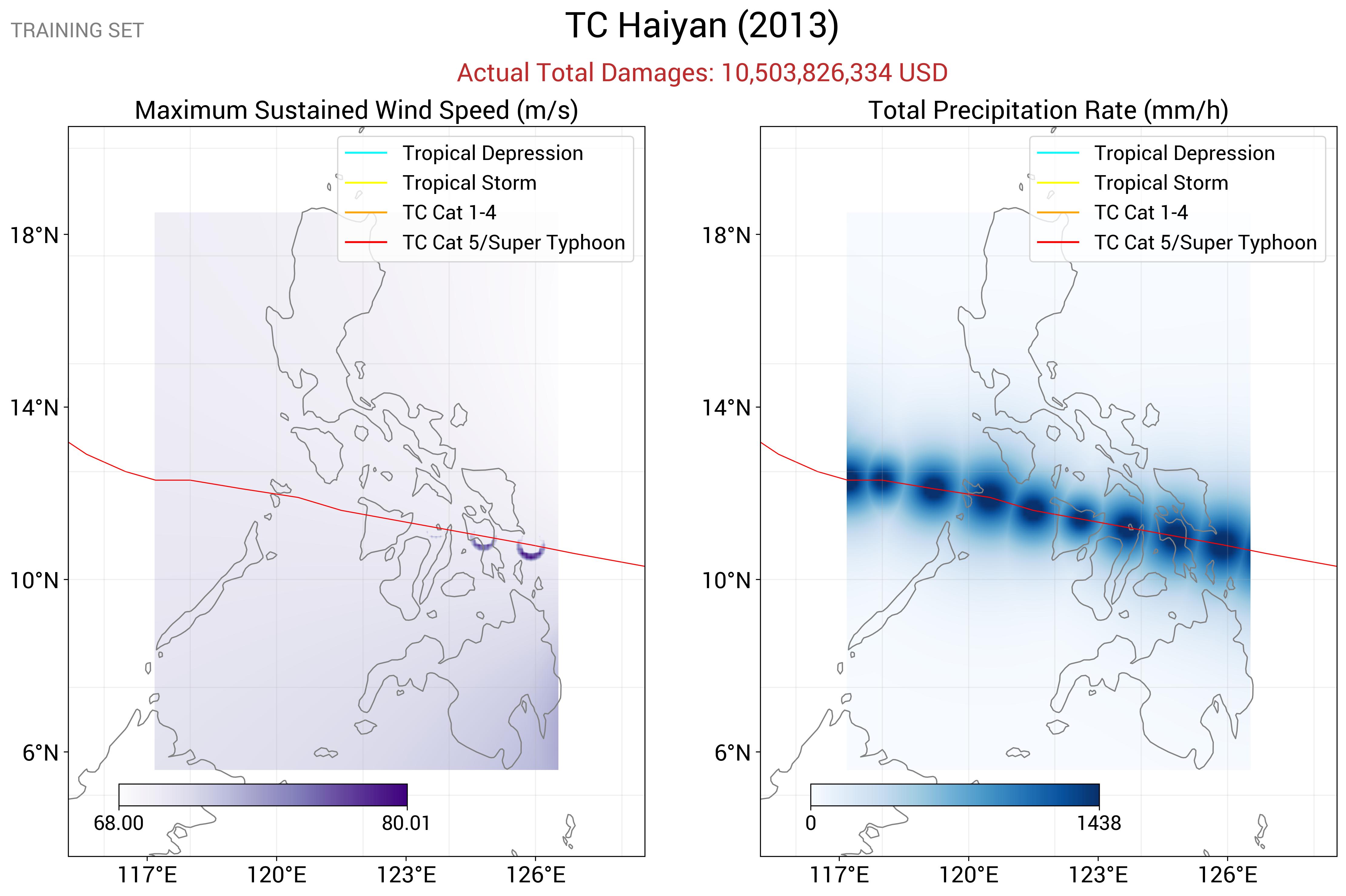} \quad \includegraphics[scale=0.23]{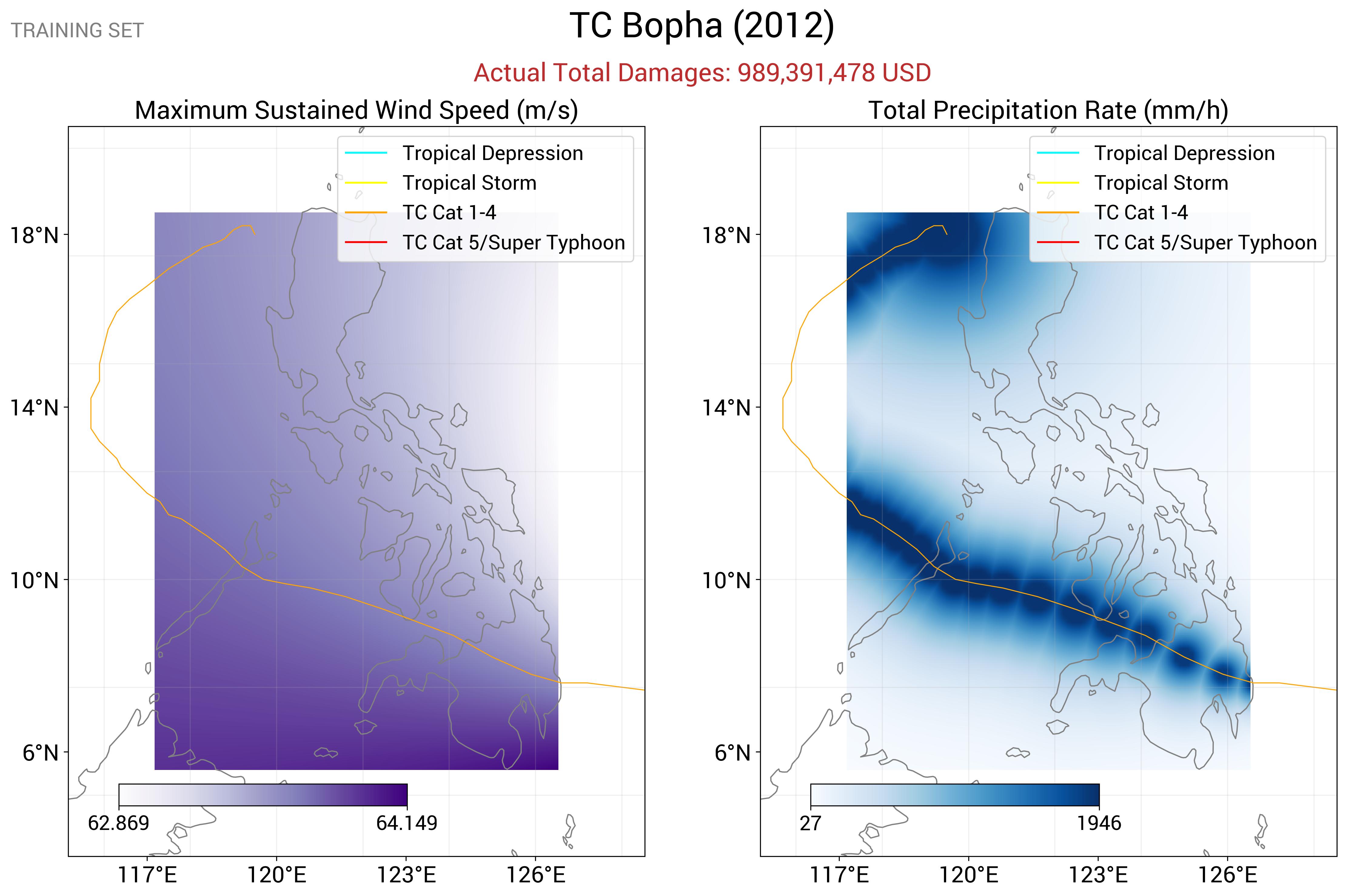} \hfill
                \includegraphics[width=0.7\textwidth]{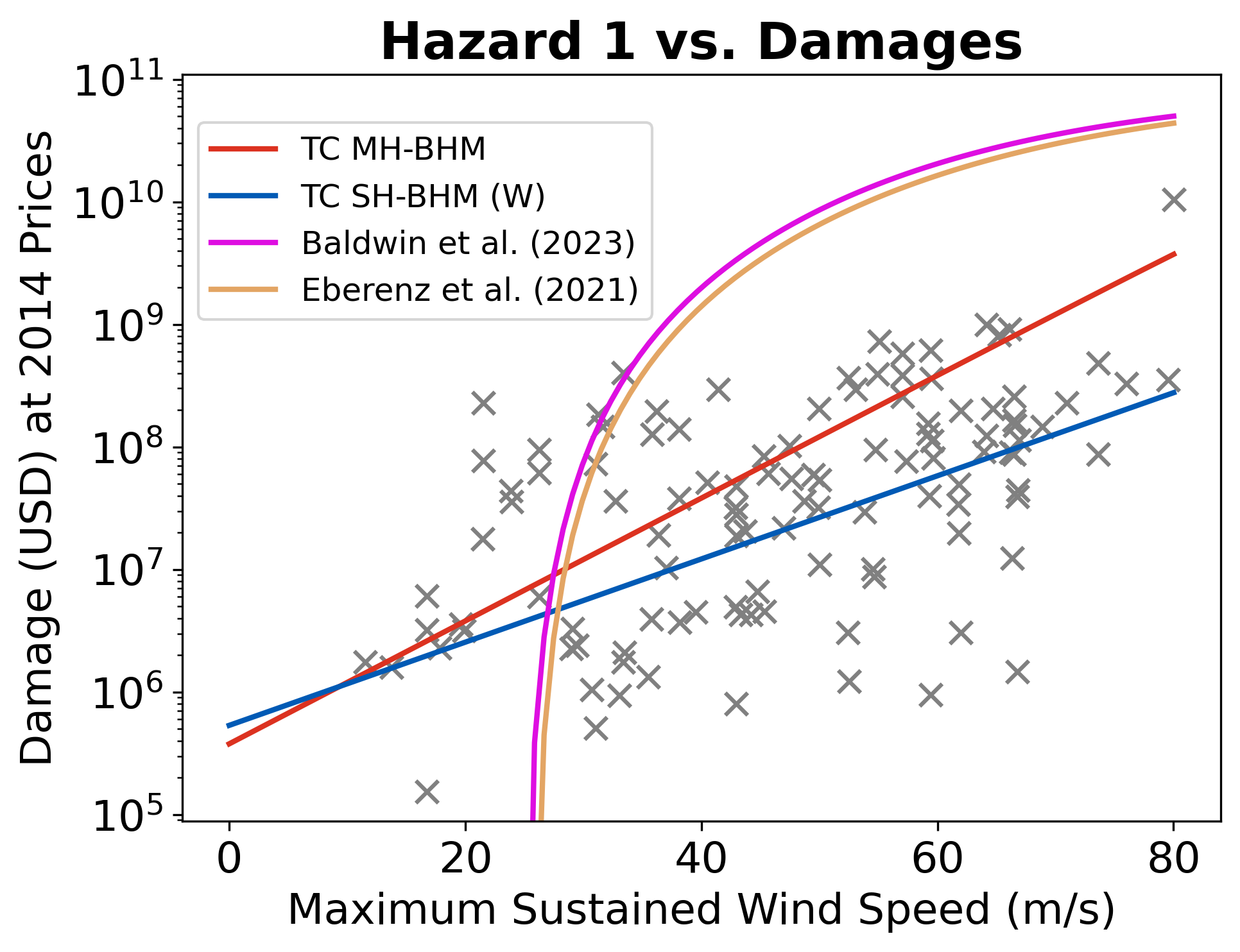}
                \includegraphics[width=0.7\textwidth]{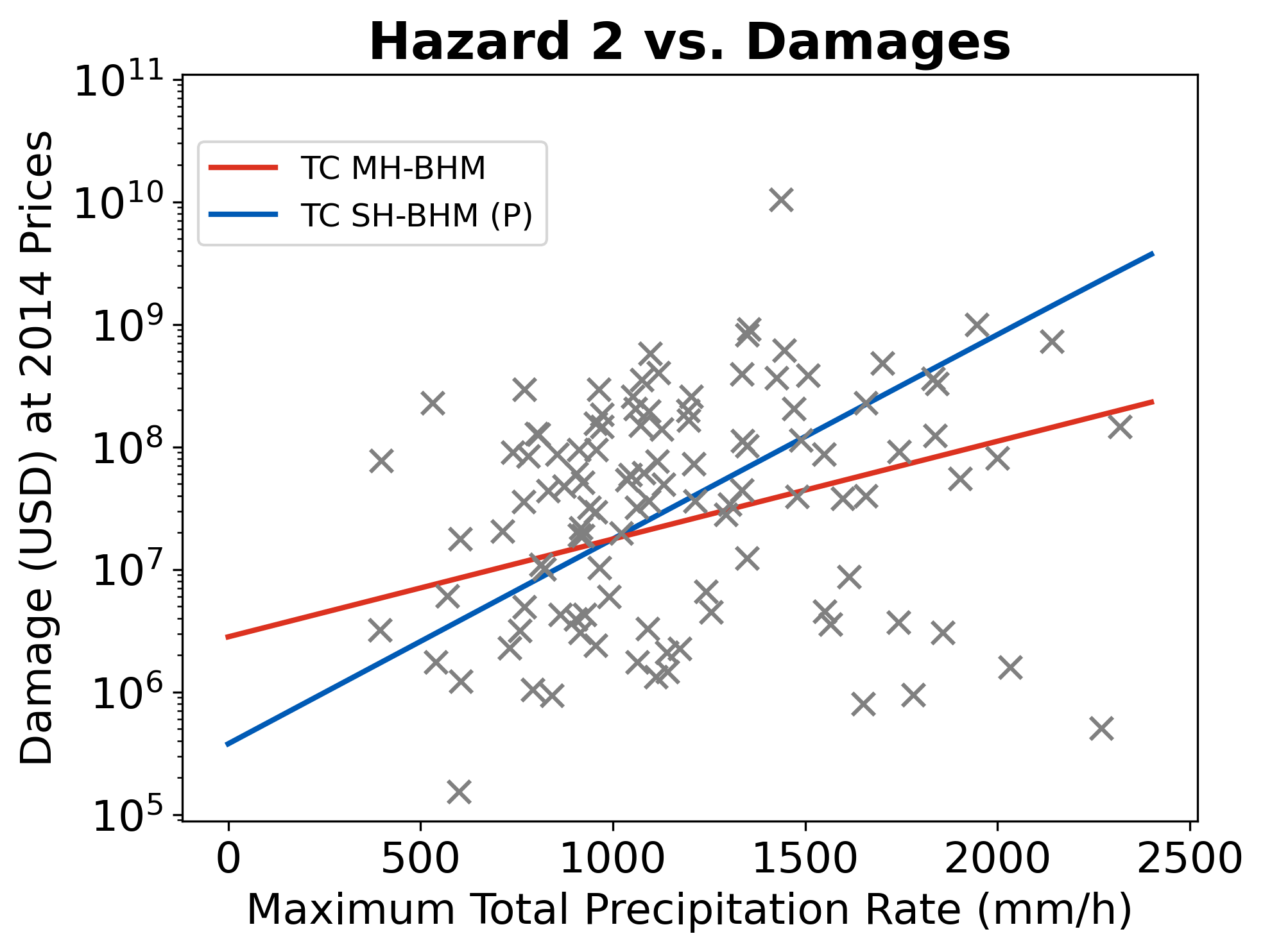}
                \caption{}
                \label{fig:tc_tracks_training}
        \end{subfigure} \quad \quad \hfill
        \begin{subfigure}[b]{0.48\textwidth}
        \centering
                \includegraphics[scale=0.23]{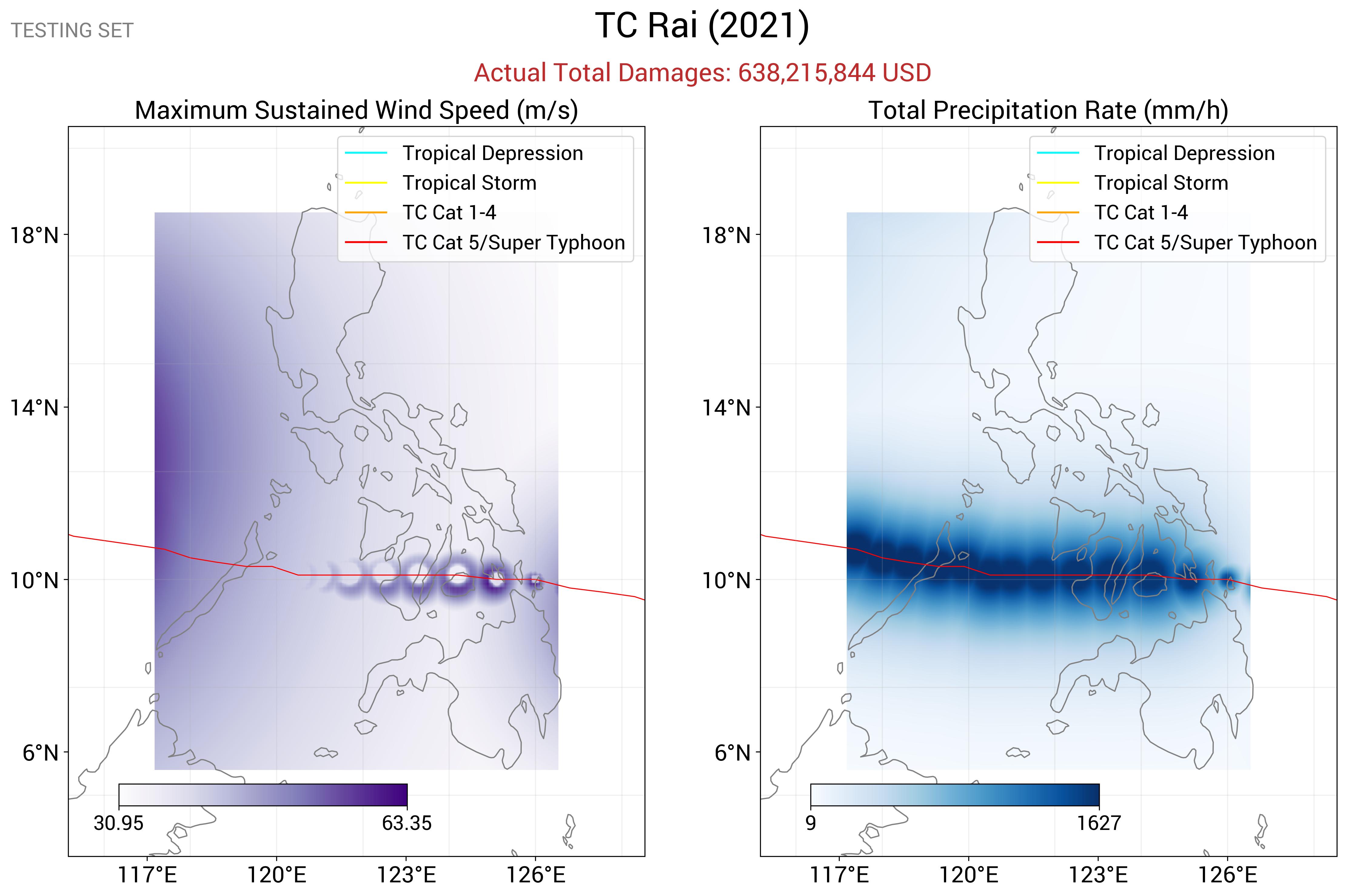} \quad \includegraphics[scale=0.23]{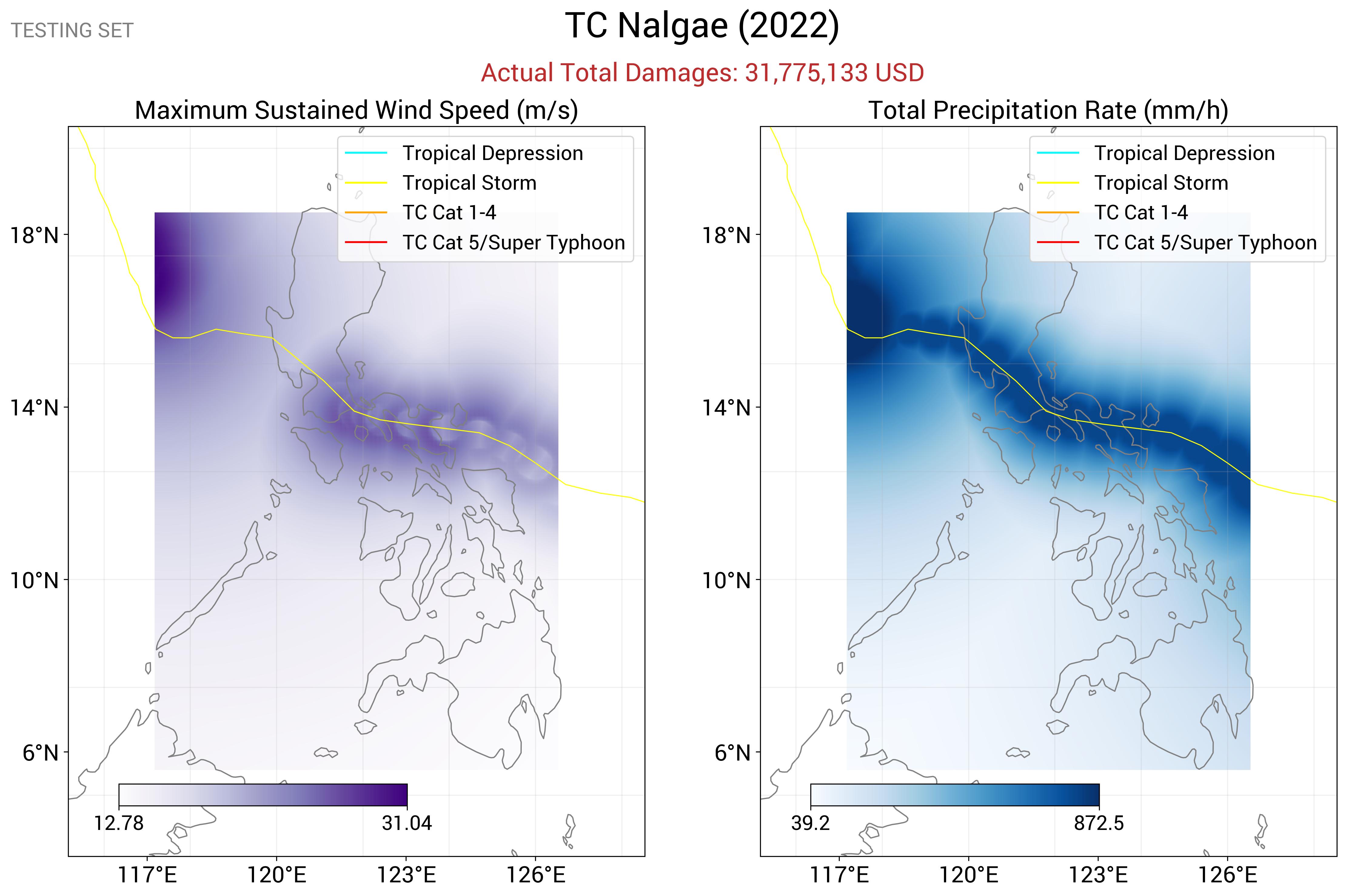} \hfill
                \includegraphics[width=0.7\textwidth]{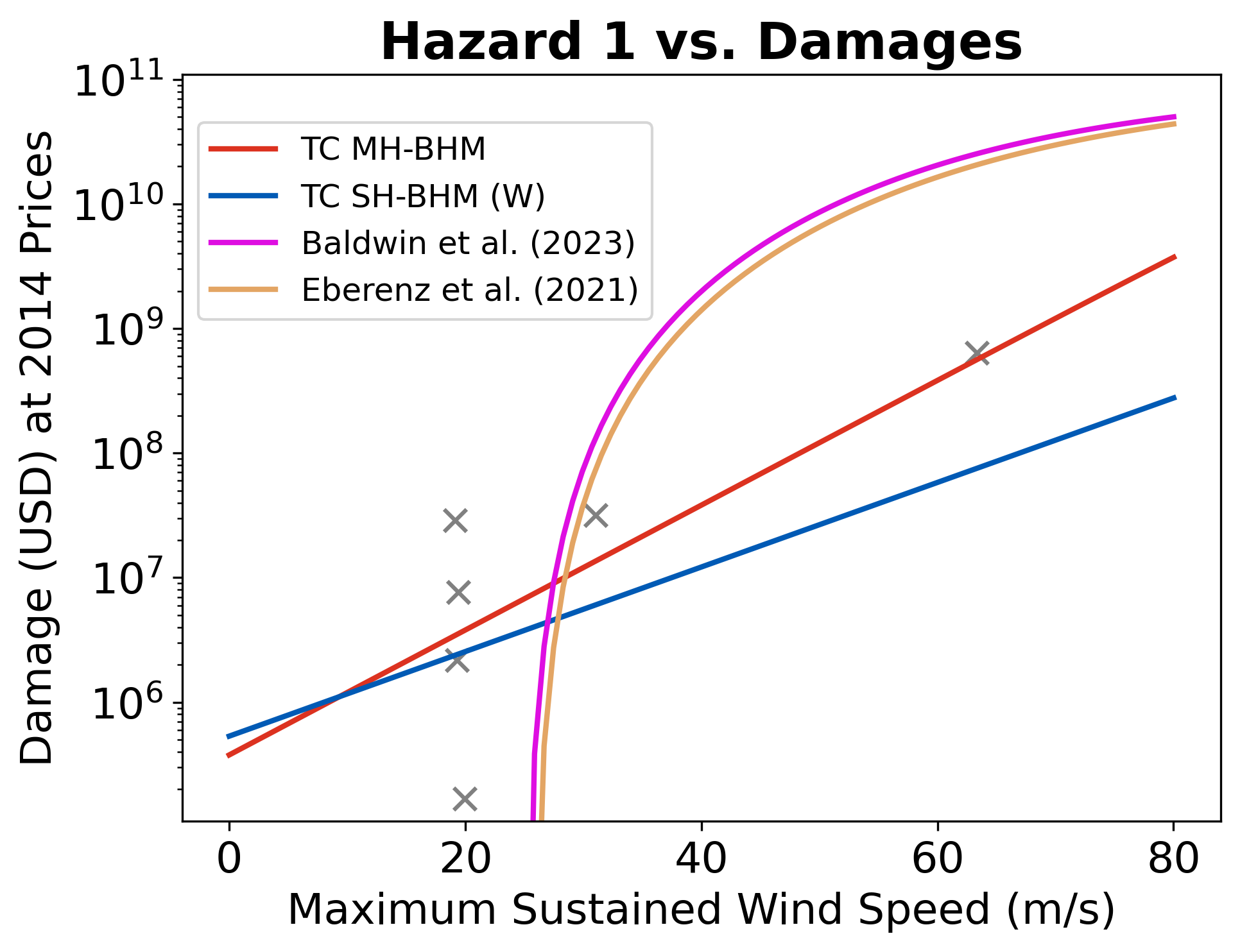} 
                \includegraphics[width=0.7\textwidth]{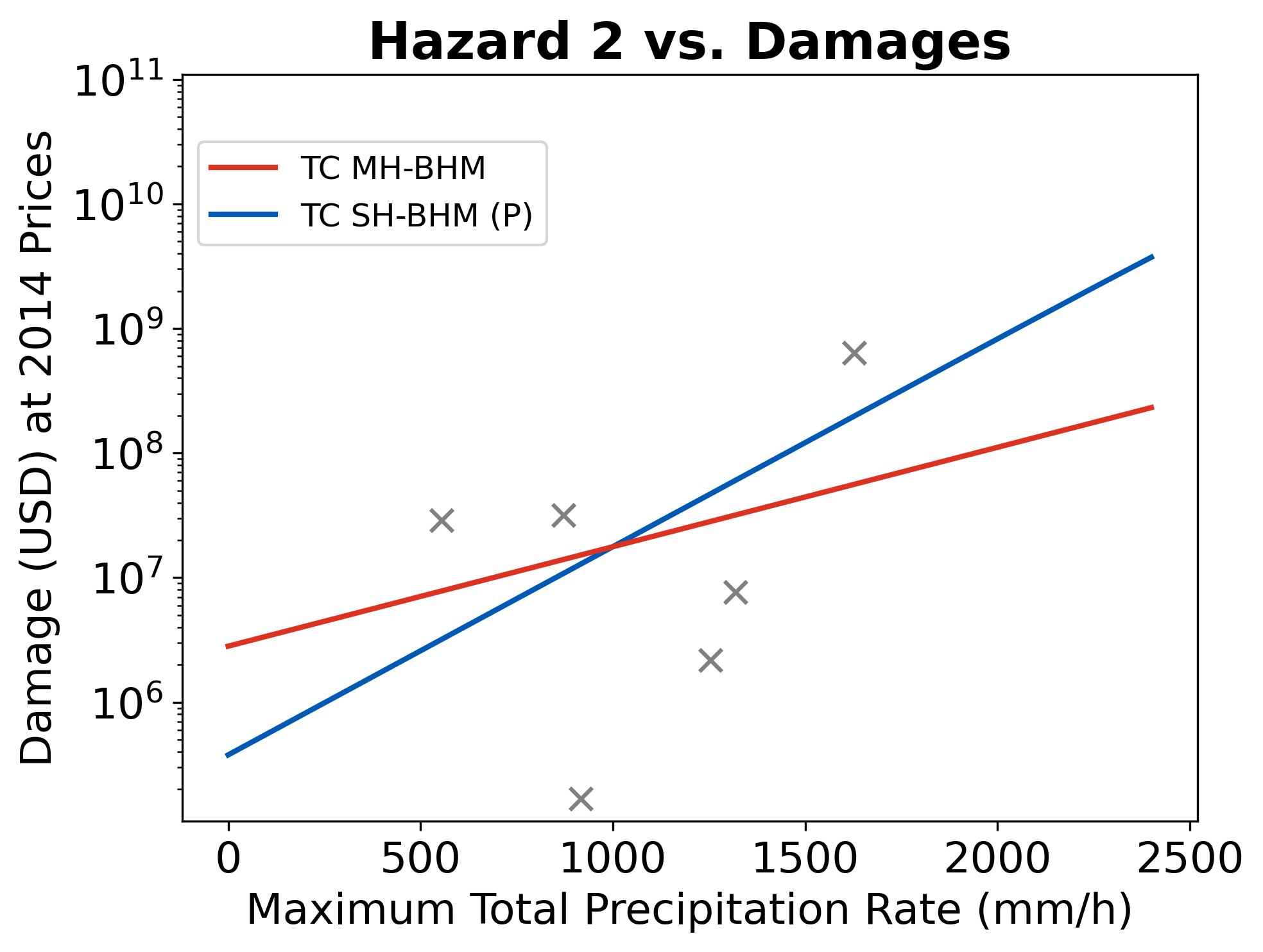} 
                \caption{}
                \label{fig:tc_tracks_testing}
        \end{subfigure}%
        
        \caption{\footnotesize (a) Training set and (b) testing set TC tracks of the two most destructive TCs in the Philippines (top two rows) and hazard intensity versus damage plots superimposed with the fitted vulnerability functions from the proposed models (MH-BHM and SH-BHM) and and those from existing models found in previous research (bottom two rows). Note that precipitation has not been considered in previous studies. Hence, no comparison curves are available for precipitation-related vulnerability.} 
\end{figure}

Figure~\ref{fig:tc_tracks_training} shows the tracks of the two most destructive TCs in the country's history, namely TC Haiyan (2013) and TC Bopha (2012), along with the hydrometeorological variables and the reported total damage. The maximum sustained wind speed and the maximum total precipitation rate for each TC included in the training set are also plotted against the reported total damage. It can be seen from the figure that as the wind speed or the precipitation increases, so does the damage. This relationship between each hazard and damage in the historical data validates the choice of the TC MH-BHM as it successfully captures this characteristic pattern. 

\subsection{Dataset Description}

The TC MH-BHM is fitted to 113 TC events represented by the TC tracks shown in Figure~\ref{fig:fig-historical-tc-tracks-ph}. The historical TC dataset was obtained from the IBTrACS (International Best Track Archive for Climate Stewardship) database version 4r01 \citep{knapp2010international, gahtan2024international}. IBTrACS provides the most comprehensive global record of historical TCs.  While the IBTrACS Philippine records extend from 1945 to the present, this work calibrates the TC MH-BHM using only 113 tropical cyclone events that occurred between 1979 and 2020. The TCs included in the training set are those that match the damage records in the International Disaster Database EM-DAT \citep{EM-DAT}. The damage values in the EM-DAT database are reported in nominal dollars corresponding to the year each event occurred, without inflation adjustments. Meanwhile, the LitPop database presents asset exposure data with all values standardized to 2014 USD prices. To eliminate the effects of economic changes over time on the actual damage values, the EM-DAT damage values are normalized to match the LitPop asset exposure data using the approach described in \cite{eberenz2021regional} and \cite{baldwin2023vulnerability}. Hence, all monetary values pertaining to the application study are in 2014 USD prices.

From the historical TC tracks, wind and precipitation fields are generated using the TC hazard model of \cite{nederhoff2024accounting}. Although the model produces wind speeds that align well with historical records, the precipitation rates from \cite{nederhoff2024accounting} significantly underestimate historical values. To address this underestimation, the model-derived precipitation rates are first standardized then rescaled using values from the Global Multi-Source Tropical Cyclone Precipitation (MSTCP) database \citep{morin2024global}. This adjustment method produces precipitation values that agree well with historical records.

The period of 1979-2020 was used to build the training set since the MSTCP database, which supplies the TC precipitation data, only provides records from 1979 to 2023. For model validation, six TCs that occurred during 2020-2022 were selected as the testing dataset, as both EM-DAT and MSTCP records were available for these events. Figure~\ref{fig:tc_tracks_testing} shows the two most destructive TCs in the testing set. 

\subsection{Results}

\begin{figure}[t]
\centering
    \includegraphics[width=\linewidth]{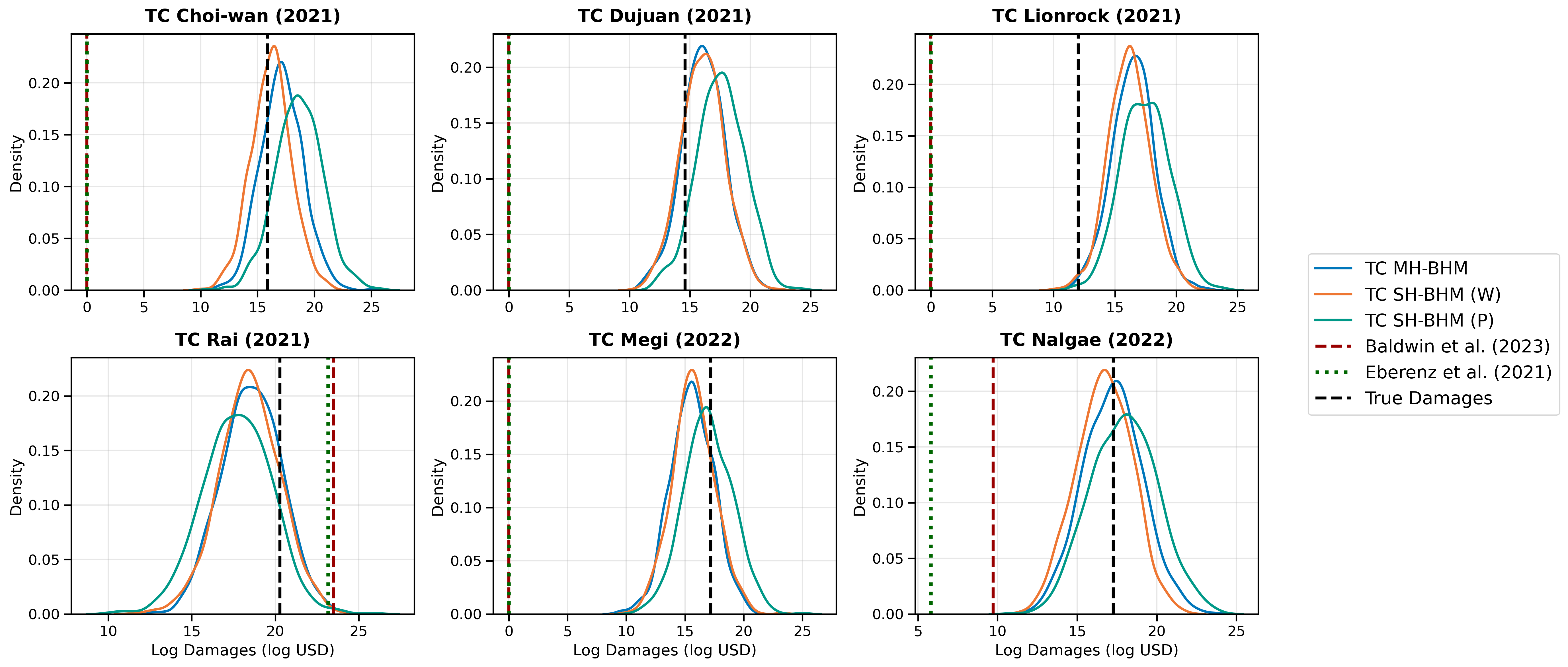}
    \caption{Out-of-sample posterior predictive distribution of damages for the six TCs in the testing set. The predictions from well-established deterministic models are also overlaid, with \cite{baldwin2023vulnerability} shown as a red dashed line and \cite{eberenz2021regional} shown as a dotted line, while the true observed damages are indicated by the black dashed line.} 
    \label{fig:fig-posterior-predictions-historical}
\end{figure}


To rigorously validate the performance of the BHMs, a comprehensive out-of-sample validation is conducted using TCs from the 2020-2022 season that were not included in the training set. Additionally, two well-established deterministic models that were specifically calibrated for the Philippines were implemented for comparison: the \cite{baldwin2023vulnerability} model, which uses the damage power vulnerability function in (\ref{eqn:emanuel_fragility_function}) with $v_{\text{thresh}} = 25$ and $v_{\text{half}} = 80$, and the \cite{eberenz2021regional} model, with $v_{\text{thresh}} = 25.7$ and $v_{\text{half}} = 84.7$.

Figure~\ref{fig:fig-posterior-predictions-historical} presents damage predictions from three competing models for six TCs that affected the Philippines in 2020-2022. The posterior density distributions from all BHM variants significantly outperform the traditional deterministic models, as evidenced by their ability to capture the true damage values within their probability distributions. Notably, the true damage values (black) consistently fall within the highest density regions of the BHM posterior distributions. The deterministic models, in contrast, often produce estimates that deviate substantially from the true damages, such that their predictions (red and green vertical lines) fall outside the central probability mass of the BHM distributions.

\begin{figure}[t]
\centering
    \includegraphics[width=\linewidth]{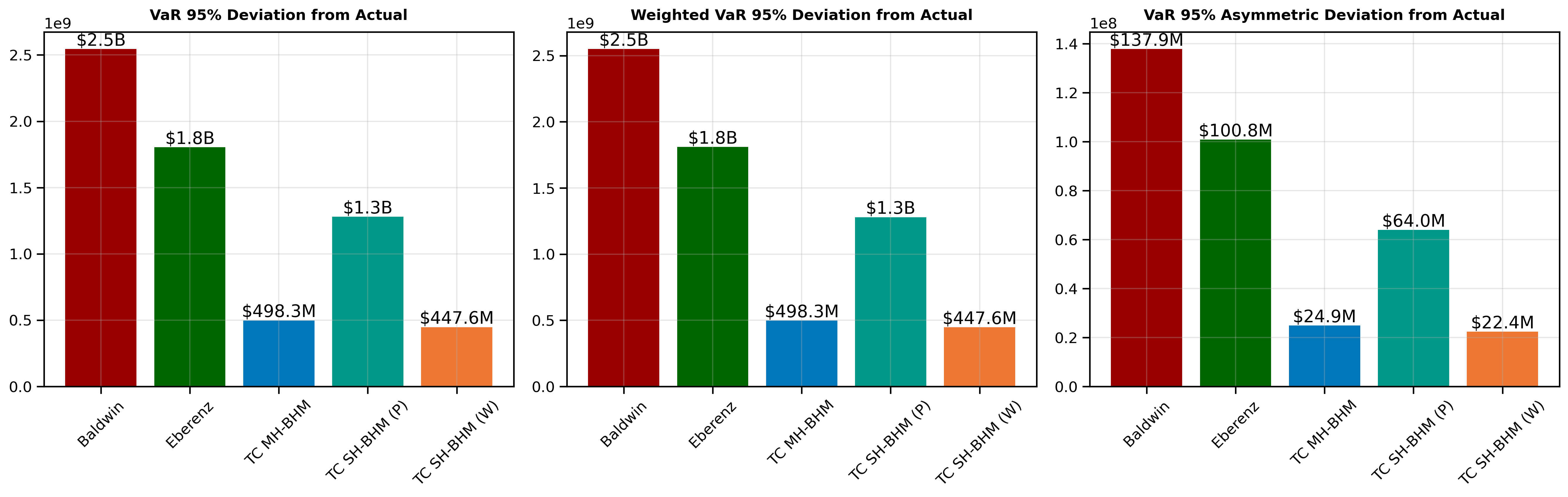}
    \caption{Risk assessment metrics across competing models (2 deterministic and 3 BHMs) for out-of-sample historical TCs. Smaller values indicating better predictive performance.} 
    \label{fig:fig-risk-metrics-comparisons-historical}
\end{figure}

Three complementary risk metrics are employed to evaluate model performance when only one actual damage value is available rather than a full distribution. First, the $\text{VaR}_{\alpha}$ deviation from actual $\text{Dev}_{\text{sym}} = \sqrt{(y_{\text{true}} - \text{VaR}_{\alpha})^2}$ provides a symmetric penalty for deviations from the true value. Second, the weighted $\text{VaR}_{\alpha}$ deviation from actual, $\text{Dev}_{\text{w}} = \sqrt{\mathbb{E}[w(e)e^2]}$, where $e = y_{\text{true}} - \text{VaR}_{\alpha}$ and $w(e)$ is a weighting function that assigns higher penalties ($w(e) > 1$) to overestimation and lower penalties ($w(e) = 1$) to underestimation. Finally, the $\text{VaR}_{\alpha}$ asymmetric deviation from actual, $\text{Dev}_{\text{asym}}(y_{\text{true}}, \text{VaR}_{\alpha}) = \max(\alpha(y_{\text{true}} - \text{VaR}_{\alpha}), (\alpha-1)(y_{\text{true}} - \text{VaR}_{\alpha}))$ asymmetrically penalizes under- and over-estimation based on the chosen confidence level $\alpha$. The last two metrics are particularly relevant for risk management where the costs of under- and over-estimation may differ. Figure~\ref{fig:fig-risk-metrics-comparisons-historical} summarizes the values of these three risk metrics. The three plots demonstrate the superior performance of BHMs compared to deterministic approaches such that BHMs consistently show smaller deviations from actual damage values, indicating more reliable predictive capabilities across different weighting schemes. 

\begin{figure}[t]
\centering
    \includegraphics[width=\linewidth]{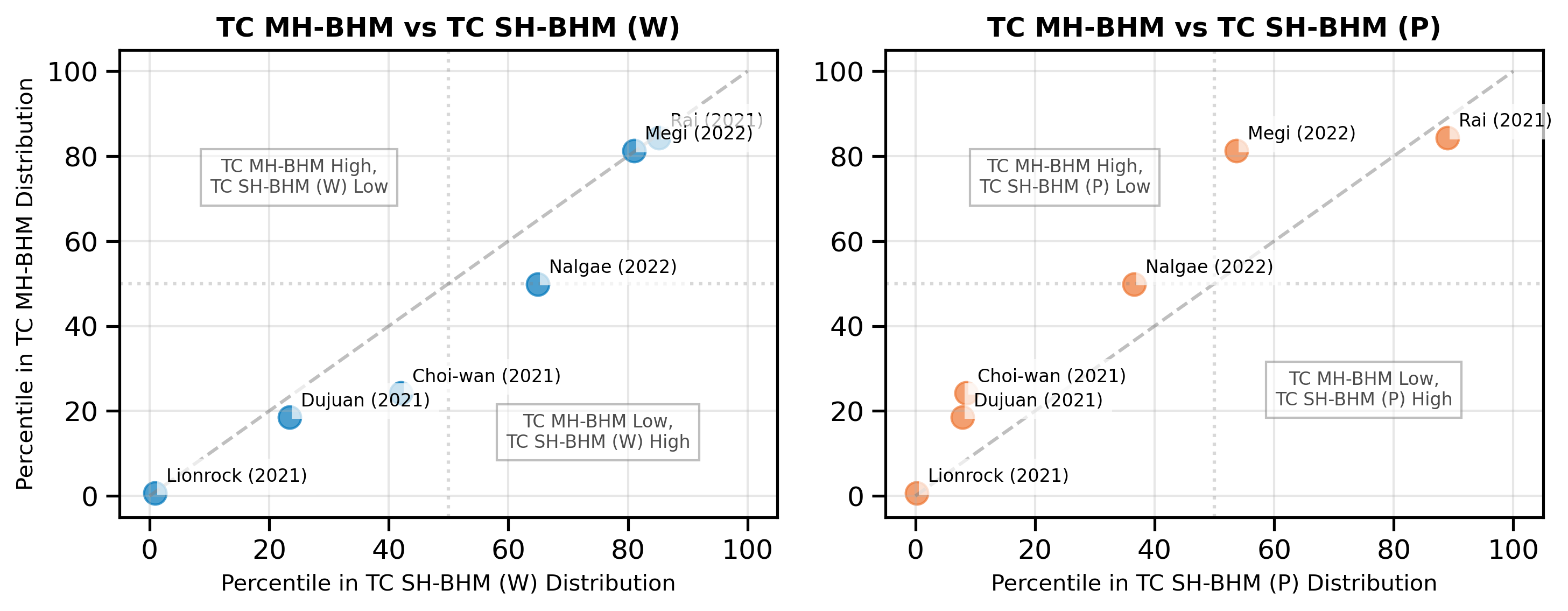}
    \caption{Percentile comparison of true damage locations  between (left) multi-hazard and single-hazard (wind only) and (right) multi-hazard and single-hazard (precipitation only). The diagonal dashed line indicates equal percentile predictions between models.} 
    \label{fig:fig-percentile-comparisons}
\end{figure}

One may also assess how well each model positions the single actual damage value within its predictive distribution rather than trying to compare full distributions of true values. Figure~\ref{fig:fig-percentile-comparisons} compares actual damage percentile locations across the BHMs. Points above the diagonal indicate TCs where TC MH-BHM predicts higher percentiles for actual damages compared to single-hazard models. TC MH-BHM demonstrates superior predictive capabilities across TCs, revealing more balanced predictions across different intensities. For high-impact events such as TC Rai (2021) and TC Megi (2022), TC MH-BHM shows strong agreement with the wind-based single-hazard model. For moderate-intensity TCs, TC MH-BHM reveals its strength by capturing complex multi-hazard dynamics, while single-hazard models tend to overemphasize isolated damage components.

\section{Conclusion} \label{sec:conclusion}

This study presents a significant methodological advancement in natural hazard risk assessment by providing the first comprehensive probabilistic treatment of the classical damage equation. The proposed Bayesian hierarchical framework systematically incorporates uncertainty across hazard, exposure, and vulnerability components while preserving the intuitive structure that practitioners value. The model demonstrates superior predictive performance through simulation studies and historical tropical cyclone data compared to traditional deterministic and single-hazard approaches. Notably, this work provides the first empirical validation of precipitation's contribution to tropical cyclone damages, highlighting the importance of considering multi-hazard interactions in risk assessment.
The MH-BHM opens several promising directions for future research. While this study focuses on the probabilistic treatment of the damage equation, future work will extend this approach to the full risk equation, incorporating additional uncertainties in hazard frequency and temporal patterns. Recognizing that total risk and subsequent damages are inherently linked to a community's ability to recover from disasters, future research will incorporate recovery dynamics into the probabilistic framework. This extension will enable modeling of both immediate damage and longer-term impacts, providing a more complete assessment of disaster risk. The model's flexibility allows for incorporating additional hazard interactions, adaptation to other natural disasters, and integration of time-varying vulnerability patterns. Immediate plans include the implementation of MH-BHM within established damage modeling platforms such as CLIMADA, enhancing existing tools with uncertainty quantification capabilities and multi-hazard functionality while maintaining their familiar interfaces. Furthermore, this framework provides a foundation for improving insurance applications, particularly in refining basis risk calculations, premium pricing, and risk transfer mechanisms through more accurate uncertainty quantification.
From a practical perspective, the publicly available implementation and datasets enable immediate application across various contexts, from insurance pricing to disaster risk reduction planning. As climate change continues to alter hazard patterns and exposure distributions evolve with urbanization, this probabilistic approach provides a robust foundation for quantifying and managing risks in an increasingly uncertain world.

\section*{Codes and Data Availability}
\sloppy The codes and data supporting the analyses and figures in this article are from the following references: (1) The LitPop dataset used for asset exposure is available at \url{https://essd.copernicus.org/articles/12/817/2020/}. (2) The Emergency Events Database (EM-DAT) used for reported damages is available online at \url{https://public.emdat.be}. (3) NOAA's International Best Track Archive for Climate Stewardship (IBTrACS v4.0) for tropical cyclone records (\url{https://www.ncei.noaa.gov/products/international-best-track-archive}). (4) Codes and replication data from Baldwin, J (2023). "Vulnerability in a Tropical Cyclone Risk Model: Philippines Case Study" (DOI:10.1175/WCAS-D-22-0049.1): \url{https://www.designsafe-ci.org/data/browser/public/designsafe.storage.published/PRJ-3803}. (5) The CLIMADA software is available on GitHub at \url{https://github.com/CLIMADA-project/climada_python}. (6) The TC emulator used in this study is available on GitHub at \url{https://github.com/Deltares-research/cht_cyclones}.

\baselineskip=20pt
\bibliographystyle{apalike}
\bibliography{main}


\newpage

\appendix
\renewcommand{\thesection}{\arabic{section}}
\renewcommand{\thesection}{\hspace{0.5em}} 
\renewcommand{\thesubsection}{\Alph{section}A.\arabic{subsection}}

\end{document}